\documentclass[fleqn,usenatbib]{mnras}

\usepackage{newtxtext,newtxmath}

\usepackage[T1]{fontenc}

\DeclareRobustCommand{\VAN}[3]{#2}
\let\VANthebibliography\thebibliography
\def\thebibliography{\DeclareRobustCommand{\VAN}[3]{##3}\VANthebibliography}

\usepackage{graphicx}	
\usepackage{multirow}
\usepackage{xfrac}      
\usepackage{gensymb}
\usepackage{inputenc}
\usepackage{tablefootnote}
\usepackage{wasysym}
\usepackage{stfloats}
\usepackage{xspace}
\usepackage{soul}

\usepackage[usenames]{color}

\newcommand{\mstar}{\mbox{$M_\star$}\xspace}
\newcommand{\msun}{\mbox{$\rm{M}_\odot$}\xspace}
\newcommand{\vsig}{\mbox{$V/\sigma$}\xspace}

\newcommand{\newh}{\mbox{{\sc \small NewHorizon}}\xspace}
\newcommand{\illustris}{\mbox{{\sc \small Illustris}}\xspace}
\newcommand{\illustristng}{\mbox{{\sc \small IllustrisTNG}}\xspace}
\newcommand{\eagle}{\mbox{{\sc \small Eagle}}\xspace}
\newcommand{\hagn}{\mbox{{\sc \small Horizon-AGN}}\xspace}
\newcommand{\tnghundred}{\mbox{{\sc \small TNG100}}\xspace}
\newcommand{\tngfifty}{\mbox{{\sc \small TNG50}}\xspace}

\definecolor{lightblue}{rgb}{0.12, 0.56, 1.0}
\definecolor{lightred}{rgb}{1.0, 0.44, 0.44}
\definecolor{Blue}{rgb}{0,0.08,0.65}
\definecolor{Red}{rgb}{0.65,0.08,0.05}
\definecolor{Green}{rgb}{0.15,0.45,0.25}

\definecolor{Pink}{rgb}{0.858, 0.188, 0.478}

\newcommand{\rd}{\mathrm{d}}

\newcommand{\ad}{a_{\mathrm{d}}}
\newcommand{\ab}{a_{\mathrm{b}}}
\newcommand{\ah}{a_{\mathrm{h}}}
\newcommand{\anm}{a_n^m}
\newcommand{\bnm}{b_n^m}
\newcommand{\cnm}{c_n^m}
\newcommand{\alm}{a_l^m}
\newcommand{\blm}{b_l^m}
\newcommand{\clm}{c_l^m}
\newcommand{\Pnm}{P_n^{|m|}}
\newcommand{\Plm}{P_l^{|m|}}

\title[Barred Galaxies in NewHorizon]{The NewHorizon Simulation - To Bar Or Not To Bar}

\author[Reddish et al.]{J. Reddish$^{1}$\thanks{E-mail: reddishjames7@gmail.com},
K. Kraljic$^{1,2}$\thanks{E-mail: katarina.kraljic@lam.fr},
M. S. Petersen$^{1,3}$, K. Tep$^{1,3}$, Y. Dubois$^{3}$,
C. Pichon$^{1,3,4}$,\newauthor S. Peirani$^{5,3}$, F. Bournaud$^{6,7}$, H. Choi$^{8}$, J. Devriendt$^{9}$, 
R. Jackson$^{8}$, G. Martin$^{11,12}$,
\newauthor M. J. Park$^{13}$, M. Volonteri$^{3}$, S. K.~Yi$^{8}$
\\
$^{1}$ Institute for Astronomy, University of Edinburgh, Royal Observatory, Edinburgh, EH9 3HJ, UK\\
$^{2}$ Aix Marseille Universit\'e, CNRS, CNES, UMR 7326, Laboratoire d’Astrophysique de Marseille, Marseille, France\\
$^{3}$ Institut d'Astrophysique de Paris, UMR 7095, CNRS, UPMC Univ. Paris VI, 98 bis boulevard Arago, 75014 Paris, France \\
$^{4}$ Korea Institute of Advanced Studies (KIAS) 85 Hoegiro, Dongdaemun-gu, Seoul, 02455, Republic of Korea\\
$^5$  Université Côte d’Azur, Observatoire de la Côte d’Azur, CNRS, Laboratoire Lagrange, Nice, France\\
$^{6}$ AIM, CEA, CNRS, Université Paris-Saclay, Université Paris Diderot, Sorbonne Paris Cité, 91191 Gif-sur-Yvette, France\\
$^{7}$ IRFU, CEA, Université Paris-Saclay, 91191 Gif-sur-Yvette, France\\
$^{8}$ Department of Astronomy and Yonsei University Observatory, Yonsei University, Seoul 03722, Republic of Korea\\
$^9$ Department of Physics, University of Oxford, Keble Road, Oxford OX1 3RH, United Kingdom\\
$^{11}$ Korea Astronomy and Space Science Institute, 776 Daedeokdae-ro, Yuseong-gu, Daejeon 34055, Republic of Korea\\
$^{12}$ Steward Observatory, University of Arizona, 933 N. Cherry Ave, Tucson, AZ 85719, USA\\
$^{13}$ Center for Astrophysics $\mid$ Harvard \& Smithsonian, 60 Garden St., Cambridge, MA 02138, USA
}

\date{Accepted XXX. Received YYY; in original form ZZZ}

\pubyear{2022}

\begin{document}
\label{firstpage}
\pagerange{\pageref{firstpage}--\pageref{lastpage}}
\maketitle

\begin{abstract}
We use the \newh simulation to study the redshift evolution of bar properties and fractions within galaxies in the stellar masses range $M_{\star} = 10^{7.25} - 10^{11.4}$ \msun over the redshift range \(z~=~0.25 - 1.3\). We select disc galaxies using stellar kinematics as a proxy for galaxy morphology. We employ two different automated bar detection methods, coupled with visual inspection, resulting in observable bar fractions of $f_{\rm bar}$ = 0.070$_{\scalebox{.5}{-0.012}}^{\scalebox{.5}{+0.018}}$ at $z\sim$ 1.3, decreasing to $f_{\rm bar}$ = 0.011$_{\scalebox{.5}{-0.003}}^{\scalebox{.5}{+0.014}}$ at $z\sim$ 0.25. 
Only one galaxy is visually confirmed as strongly barred in our sample. This bar is hosted by the most massive disk and only survives from $z=1.3$ down to $z=0.7$. Such a low bar fraction, in particular amongst Milky Way-like progenitors, highlights a missing bars problem, shared by literally all cosmological simulations with spatial resolution $<$100 pc to date.
The analysis of linear growth rates, rotation curves and derived summary statistics of the stellar, gas and dark matter components suggest that galaxies with stellar masses below $10^{9.5}-10^{10}$ \msun in \newh appear to be too dominated by dark matter relative to stellar content to form a bar, while more massive galaxies typically have formed large bulges that prevent  bar persistence at low redshift.  This investigation confirms that the evolution of the bar fraction puts stringent constraints on the assembly history of baryons and dark matter onto galaxies. 
\end{abstract}

\section{Introduction}
\label{Intro}

Stellar bars are not only visually striking, they also play an important role in the evolution of disc galaxies, allowing them to redistribute their angular momentum \citep{Lynden-Bell72,Lynden-bell1979,Tremaine_Weinberg_1984,Weinberg_1985,Athanassoula_Sellwood_1986}. Bars are commonly thought to form spontaneously in stellar discs that are sufficiently massive and dynamically cold to be gravitationally unstable \citep[e.g.][]{Toomre_1963,Combes_Sanders_1981,Athanassoula_Sellwood_1986,Combes_Elmegreen_1993}.
Once formed, simulations show that bars can evolve through angular momentum exchange with both the dark matter haloes \citep[e.g.][]{Debattista_Sellwood_2000}, and with the stellar and gaseous discs \citep{Bournaud_combes_2002,Bournaud_2005}. Bars may be amplified through the transfer of angular momentum to the dark matter halo \citep[e.g.][]{Athanassoula_2008,Kormendy_2013,Petersen_2016,Petersen_2019_2}. However, they can also be weakened or destroyed if they gain too much angular momentum from infalling gas \citep{Bournaud_2005} or from excessive growth of the central mass concentration of a galaxy -- specifically, simulations suggest that short bars of $\sim$ 1 kpc may be destroyed by the growth of a galaxies central concentration \citep[e.g][]{Hasan_Norman_1990,Hasan_1993,Shen_Shellwood_2004} henceforth impacting significantly on the growth of a central bulge without the involvement of a galaxy merger \citep[e.g.][]{Du_2017,Guo_2020}. With these statements, it is important to highlight the caveat made by many of these referenced works whereby these scenarios are not yet supported by observational evidence \cite[see][and references therein for further discussions]{Shen_Shellwood_2004,Athanassoula_2005}.

Simulations have shown that bars may also be affected by the conditions of their host galaxy. They may be reinforced or reformed given sufficient accretion of external gas onto the disc \citep{Bournaud_combes_2002} as well as also being destroyed or (re-)formed as a result of environmental factors such as tidal interactions 
or galaxy mergers \citep[e.g.][]{Hohl_1971,Noguchi_1987,Gerin_1990,Berentzen_2004,Peirani_2009,Moetazedian_2017,Cavanagh_2020,Zhou_2020}. 
In its host galaxy, the presence of a bar might also effect the formation of a pseudo-bulge at the galaxy’s centre \citep[e.g.][]{Pfenniger_Norman_1990,Kormendy_Kennicutt_2004,Athanassoula_2012,Kormendy_2013,Lin_2020}. However, we note that for these scenarios, observational evidence is as yet unclear. 
Several observational studies \citep[e.g.][]{Sersic_Pastoriza_1965,Martinet_Friedli_1997,Sheth_2005,Ellison_2011,Lin_2020} suggest that bars might enhance the star formation at the centre of the galaxy, but may also play a role in the cessation of star formation \citep[e.g.][]{Masters_2012,Wang_2012,Fraser-McKelvie_2020}. Inflowing gas driven by a bar has also long been implicated in the fuelling of the central active galactic nuclei (AGN) \citep[e.g.][]{Knapen_2000,Oh_2012,Galloway_2015,Alonso_2018}, although again, compelling observational evidence is still missing  \citep[e.g.][]{Regan_Mulchaey_1997,Cisternas_2013,Cheung_2015,Cisternas_2015,Oh_2012}.
It therefore seems apparent that bars play a key role in the evolution of disc galaxies \citep[see \citealt{Kormendy_Kennicutt_2004,Kormendy_2013}, however, for an opposing view, see][and references therein]{van_der_bergh_2011}.

In spite of the existence of increasingly constrained observational measurements of the frequency of bars in the local Universe \cite[e.g.][]{Eskridge_2000,Whyte_2002,Laurikainen_2004,Menendez-Delmestre2007,Marinova_Jogee_2007,Barazza_2008,Aguerri_2009,Sheth_2008,masters_2011,Masters_2012,Oh_2012,Melvin_2014,Diaz_2016}, and  a number of galaxy simulations routinely forming bars \cite[e.g.][]{early_bar_sims_1,early_bar_sims_2,Athanassoula_2008,Athanassoula_2012, Eagle_Bars_2017,Zhou_2020,Rosas_Guevara_2020,Rosas-Guevara_2021}, their origin and details on their life-cycle are still not fully understood. Studying the frequency of bars and the evolution through redshift as a function of stellar mass provides an opportunity to learn about the formation and persistence of bars.

In the local Universe, the fraction of barred galaxies is quite well established.
Both optical and near infrared observations reveal that roughly 2/3 (or 1/3 if only so-called “strong” bars are counted) of nearby spiral galaxies host a bar \citep[e.g.][]{Eskridge_2000,Whyte_2002,Laurikainen_2004,Menendez-Delmestre2007,Barazza_2008,Sheth_2008,Aguerri_2009,Nair_2010a,masters_2011,Masters_2012,Melvin_2014,Diaz_2016}. However, depending on the bar classification method, the strength of the bars observed, which wave-bands they have been observed in and the process of disc galaxy selection, this value can vary.  By analysing the sample of spiral galaxies in the Spitzer Survey of Stellar Structure in Galaxies (S$^4$G) in the nearby Universe ($z<0.01$), \citet{Erwin_2018} found that the fraction of barred galaxies, $f_{\rm bar}$, reaches a maximum of $\sim$ 70\% at galaxy stellar mass $\sim $ 10$^{9.7}$ \msun and then declines for both higher and lower masses.  
These measurements are in a good agreement with previous S$^4$G based study of \cite{Diaz_2016} once the same galaxy selection is applied \citep[see][for a more detailed discussion]{Erwin_2018}.   
Such findings are however in contrast to most Sloan Digital Sky Survey (SDSS) based studies \citep[e.g.][]{Masters_2012,Oh_2012,Melvin_2014} which report a strongly increasing $f_{\rm bar}$ with stellar mass (for \msun $> 10^{10}$ \msun typically). As suggested by \cite{Erwin_2018} - and demonstrated via a simulated cut in bar size - this discrepancy may be the consequence of the significant challenge that SDSS-based studies have in picking out smaller bars in galaxies with stellar mass $< 10^{10}\msun$; the ensuing underestimate of the bar fraction at these stellar masses results in the incapability of tracing a peak in bar fraction at $\sim10^{9.7}\msun$.

The bar fraction is less well constrained out to high redshifts due to a lack of spatial resolution necessary to resolve bars. 
Early conflicting results on whether the bar fraction declines towards higher redshifts \citep[e.g.][]{Abraham_1999} or stays roughly constant up to $z \sim 1$ \citep[e.g.][]{Jogee_2004,Elmegreen_2004,Sheth_2003} are today interpreted as being due to the selection bias in shallow or low resolution data targeting more massive or large systems \citep{Sheth_2008}. 
Thanks to the high resolution deep optical and near-infrared data, it is today established that the fraction of barred galaxies decreases with increasing redshift \citep[e.g.][]{Cameron_2010,Melvin_2014,Simmons_2014}. 
These high redshift studies focus primarily on stellar masses above 10$^{10}$ \msun finding increasing bar fractions with stellar masses at all redshifts up to $z \sim 1$ \citep[e.g.][]{Melvin_2014}.

Numerical simulations provide a powerful tool in studying the formation and evolution of bars. Large-scale hydrodynamical simulations, e.g. \illustris \citep{Illustris}, \illustristng \citep{Illustris-TGN}, \eagle \citep{Eagle} or \hagn \citep{Dubois_2016}, all reproduce a morphological mix of galaxies that is in a good agreement with the well established observational results. These simulations have a typical spatial resolution of $\sim$~1~kpc which is insufficient for properly resolving the scale height of galactic discs and therefore for studying bars. 
However, some recent studies attempted to quantify properties of barred galaxies in a cosmological hydrodynamical realisations of the IllustrisTNG suite - the \tnghundred simulation \citep{Rosas_Guevara_2020,Zhou_2020,Zhao_2020_Barred_IllustrisTNG} and the \tngfifty simulation \citep{Rosas-Guevara_2021} - in \illustris \citep{Peschken2019} and in \eagle \citep{Eagle_Bars_2017}. They principally focused on disc galaxies with stellar masses \mstar $\geq$ 10$^{10-10.5}$ \msun mostly at $z=0$, but some extended the analysis to higher redshifts \citep{Peschken2019,Zhao_2020_Barred_IllustrisTNG,Rosas-Guevara_2021}.
Overall, while at $z=0$ these cosmological simulations produce a fraction of barred galaxies from about 20\% \citep[in \illustris;][]{Peschken2019} and 30\% \citep[in \tngfifty;][]{Rosas-Guevara_2021} to about 40\%-60\% \citep[in \eagle;][and in \tnghundred; \citealt{Rosas_Guevara_2020,Zhao_2020_Barred_IllustrisTNG}]{Eagle_Bars_2017},
they generally overproduce bars at $z=0$ at high masses (\mstar $\gtrsim$ 10$^{10.5}$ \msun) and tend to suppress their formation at low masses (in the stellar mass range \mstar $\sim$ $10^{10-10.5}$ \msun) compared to observations \citep[][]{Erwin_2018}. These simulations also fail to reproduce the declining bar fraction with increasing redshift, seen in most observations.  

Idealised simulations reach high enough resolutions to study bars in great detail \cite[e.g.][]{Debattista_Sellwood_2000,Athanassoula_2002,Berentzen_2004,Athanassoula_2012}, however, in order to study the cosmic evolution of bar fraction, galaxies need to be modelled self-consistently, within the cosmological context, and with sufficiently high resolution, e.g. by using  zoom-in simulations \cite[e.g.][]{Kraljic_2012,Scannapieco_Athanassoula_2012,Zana_2018,Zana_2019,BlazquezCalero_2020}. Sufficient resolution in recent cosmological simulations have allowed study into the evolution of detailed structure in disk galaxies \cite[e.g.][]{Guedes_2011,Stinson_2013,Marinacci_2014,Okamoto_2015}. However, obtaining a statistically significant sample of galaxies for a meaningful study of bar fractions remains difficult. One notable exception is the work of \citet{Kraljic_2012} analysing a sample of cosmological zoom-in simulations of 33 Milky Way-mass galaxies. They found that the fraction of observable bars among spiral galaxies drops from about 80\% at \(z = 0\) to about 15\% at  $z\sim $ 1 and to almost zero at $z\sim $ 2. 
The emergence of bars in the studied mass range was established to redshift $z\sim$ 0.8 - 1. 

In this paper we present a study of the bars in the recent high-resolution cosmological simulation \newh \citep{Dubois_2020}, modelling the population of galaxies with low-to-intermediate stellar masses ($10^{7.25} \msun \lesssim \mstar \lesssim 10^{11.4} \msun$) in an average density region probing field and group environments. This mass range is wider than previously studied in simulations \cite[e.g.][]{Athanassoula_2008,Athanassoula_2012,Rosas_Guevara_2020,Zhou_2020}, which typically focus on Milky Way-mass galaxies (\mstar $\sim 10^{10 - 11} \msun$). We analyse the structure of a large sample of galaxies extracted from the simulation from redshift \(z=1.3\) down to redshift \(z=0.25\) in order to constrain the evolution of the bar fraction over this redshift range. Studying galaxies across both mass and time is crucial for accurately characterising the presence (or lack) of bars. Similar exercise for future observational data will also help create a more accurate census for bar fraction, in particular toward lower masses and higher redshifts.

This work is a comprehensive look at the bar fraction across a wide range in stellar mass and across a significant fraction of cosmic time. We implement two different automated bar detection methods so as to provide a more robust measurement of the fraction of barred galaxies. These results are then discussed on an object-by-object basis and as populations, using two distinct summary statistics.
The paper is organised as follows. In Section~\ref{Method}, we present a description of the \newh simulation along with the process of galaxy sample selection and then the bar identification methods applied are presented in Section~\ref{bar_analysis}. Section~\ref{Initial_results} contains the results: the evolution of the bar fraction from \(z=1.3 \) to \(z=0.25\). We perform a cursory dynamical analysis of individual galaxies in Section~\ref{sec:lowbarfractions}, and present two statistical analyses motivating the observed lack of bars. We note biases in our methods and the impact of galaxy substructure in Section \ref{sec:technicalities}. We then discuss the results in the context of observations and other numerical simulations in Section~\ref{Discussion}, summarising and concluding in Section~\ref{Summary}.

\section{Simulation and Galaxy Selection}
\label{Method}
\subsection{New Horizon simulation}

The simulation sample studied in this paper is taken from the high-resolution hydrodynamical cosmological zoom-in simulation \newh, a detailed description of which is presented in \citet{Dubois_2020}. 
The parent \hagn simulation \citep{Dubois_2014}, with box size of 142 Mpc, successfully reproduces galaxies in the various cosmic environments with a reasonable morphological mix \citep[e.g.][]{Dubois_2016,Kaviraj_2017,Martin_2018}, with thin and thick disks with scale heights and luminosity ratios as observed \citep{Park_2021}. However, due to its limited spatial and stellar mass resolution ($\sim$ 1 kpc and $2\times10^6\msun$, respectively), it is not well suited to study detailed structures of galaxies such as bars. 
The \newh simulation resimulates a sphere with a radius of 10~Mpc from \hagn with a dark matter (DM) mass resolution of $1.2 \times 10^{6} \msun$ (compared to $8\times 10^{7}$ \msun in \hagn), a stellar mass resolution of $1.3 \times 10^4$ \msun (compared to $2\times 10^{6}$ \msun in \hagn) and a maximum spatial resolution of 34 pc (compared to 1 kpc in \hagn). The exquisite resolution makes this simulation ideal for studying inner sub-structures within galaxies such as bars. 

\newh adopts a $\Lambda$CDM cosmology compatible with the WMAP-7 data~\citep{komatsuetal11} with cosmological parameters: Hubble constant $H_{0}$ = 70.4 km s$^{-1}$ Mpc$^{-1}$, total mass density $\Omega_{m}$ = 0.272, total baryon density $\Omega_{b}$ = 0.0455, dark energy density $\Omega_{\Lambda}$ = 0.728, amplitude of power spectrum $\sigma_{8}$ = 0.809, and power spectral index $n_{s}$ = 0.967.

\cite{Dubois_2020} describe in detail the simulation techniques employed and present several key fundamental properties of simulated galaxies including the galaxy mass function, the cosmic star formation rate, the galaxy sizes, their stellar kinematics, morphology and their evolution with redshift that are broadly in agreement with the literature.

\subsection{Galaxy Selection}
\label{Galaxy_Selection}
Galaxies are identified and extracted from the \newh simulation using either the AdaptaHOP \citep{Aubert_2004} or the HOP \citep{HOP_Paper_1998} galaxy identification algorithm. The latter keeps all the substructures (i.e. star forming clumps and connected satellites) in a galaxy whereas the former removes the majority of these substructures. The AdaptaHOP selected galaxies will be the primary focus of this analysis, however, HOP galaxies will also be studied and discussed in Section \ref{impact_of_substructures}.

To study bars across a wide range of masses, we select a minimum galaxy mass of $\mstar \geq 10^{7.25} \msun$ \citep[see also][]{Dubois_2020} 
resulting in 525, 479, 416, 381 and 299 galaxies above this mass cut at redshifts 1.3, 1.0, 0.7, 0.5 and 0.25, respectively\footnote{The mass cut corresponds to a minimum of 2000 stellar particles.}. A discussion on galaxy and halo identification, as well as the number of galaxies for different stellar mass thresholds identified, can be found in Section 2.6 of \citet{Dubois_2020}.

The bar fraction is defined in terms of spiral galaxies. Therefore, disc-dominated (spiral) galaxies must be distinguished from spheroidal-dominated (early-type) galaxies. We do this using stellar kinematics of galaxies as a proxy to infer a galaxy's morphology. We first find the spin of a galaxy by measuring its angular momentum vector from the velocity vectors of its stars. This spin vector defines the orientation of the z-axis cylindrical coordinate system to which the Cartesian coordinates of each stellar particle are transformed. The rotational velocity $V$ of the galaxy is the average of the tangential velocity component. The velocity dispersion $\sigma = \sqrt{(\sigma_{r}^{2} + \sigma_{t}^{2} + \sigma_{z}^{2})/3}$, where $\sigma _{r,t,z}^{2} = \left \langle V_{r,t,z}^{2} \right \rangle - \left \langle V_{r,t,z}\right \rangle ^{2}$,
with the subscripts $r,t,z$ indicating the radial, tangential and vertical components, respectively.
The ratio \vsig then provides an insight into the morphology of a galaxy. 
We use this parameter to select a sample of disc (rotation-dominated) galaxies, by adopting \vsig~>~0.5, consistently with the criterion used in \cite{Dubois_2020}\footnote{A similar cut also reproduces the observed disc/elliptical fraction trends in the \hagn simulation \citep[see][]{Martin_2018}.}.
Our conclusions do not change when a higher threshold is used.
The kinematics are computed on stellar particles, AdaptaHOP or HOP, depending on the catalogue used, extracted within two effective radii ($R_{\rm eff}$; obtained from the half-mass radius of a galaxy, see \citeauthor{Dubois_2020} \citeyear{Dubois_2020} for more details).
The resulting fractions of spiral galaxies at all analysed redshfits are shown in Table~\ref{tab:Bar_Fractions}.

\section{Bar Analysis}
\label{bar_analysis}

Visual analysis of the surface density of galaxies has long been standard practice in observational studies for both morphological classification of galaxies and the identification of bars, spirals and other stellar structures \cite[see][for both seminal works and for more recent innovations]{de_Vaucouleurs_1963,Sandage_2005, Nair_2010a,Buta_2010,masters_2011}. 
However, visual inspection may suffer from ambiguity as to what constitutes a bar feature. Therefore, algorithmic surface density decompositions, such as Fourier analysis, are a natural step to ensure reproducibility. 

We will employ here two different algorithmic bar detection methods relying on Fourier analysis. The first method  (presented in Section \ref{azimuthal_analysis}) is based on an azimuthal spectral analysis of the stellar surface density of the galaxy \citep[][]{Kraljic_2012}, while the second  (described in Section \ref{Harmonic_Decomposition}) relies on a harmonic decomposition of the velocity profile of the galaxy \citep[][]{Petersen_2019_1}. We will see that comparing the two methods alongside inspecting galaxies visually produces a more robust measurement of the bar fraction than either method alone.  

\subsection{Surface Density Harmonic Decomposition}
\label{azimuthal_analysis}

The first bar detection method is presented in detail in \citet{Kraljic_2012} and involves the azimuthal spectral analysis of the surface density of galaxies. In brief, we compute the stellar surface density of the galaxy, perform a fast Fourier transform (FFT) and analyse the Fourier components. 

We compute the stellar surface density of the galaxy in the 
face-on projection (with the spin axis of the entire stellar content of the galaxy being used to define the corresponding line of sight). We take the size of the radial bins to be 100 pc so as to incorporate a few resolution elements, reducing the noise when calculating the Fourier phase. We then decompose the stellar surface density into Fourier components in the form
\begin{equation}
\label{Fourier_Decomposition_Eq}
    \Sigma (r, \theta ) =  \Sigma_{0}(r) +\sum_{m} A_{m}(r)\cos (m\theta - \Phi_{m}(r)),
\end{equation}
where $\Sigma (r, \theta )$ is the stellar surface density, $\theta$ is the azimuthal angle, and $r$ is the radial distance. $A_{m}$ and $\Phi_{m}$ are the associated Fourier amplitude and phase, respectively. $\Sigma_{0}(r)$ is the azimuthally averaged profile of the stellar surface density. A typical signature of a bar being present in a galaxy is the prominence of even Fourier components, especially the $m = 2$ mode. This bar identification process studies the Fourier phase $\Phi_{2}(r)$, which is constant with radius within the 'bar region' when a bar is present (as opposed to the spiral arm of a galaxy whereby $\Phi_{m}$ would vary linearly with radius). 
For a bar to be present, $\Phi_{2}(r)$ is allowed to vary over a given range within $\pm 5^{\circ}$ around the median value \citep[see][]{Kraljic_2012}.

To ignore variations caused by central structure, the bar search started at a radius of 0.5 kpc \cite[the inclusion of sub-kpc bars is intentional as observations point to their important contribution to the overall population; e.g.][although we note that our results do not change if we instead start at 1 kpc]{Erwin_2018}.
This search for the start of the bar region stopped at a radius of 2 kpc if no bar was found, as typically no bar identified visually starts its region of constant $\Phi_{2}(r)$ at larger radii \cite[e.g.][]{masters_2011,Melvin_2014,Diaz_2016}. The cutoff in the bar length is determined mainly by the resolution. Spiral arms typically have variations in $\Phi_{2}$ of a few tens of degrees over a distance of a few kpc but can appear constant over a short distance. Because of this, it was decided that the bar region had to cover at least 1000 pc in order to exclude the structure of any spiral arms. 
We show in Section~\ref{bar_properties} that relaxing this constraint on the minimum length of the bar does not impact the bar fractions.

\begin{figure*}
    \centering
    \includegraphics[width=0.8\textwidth]{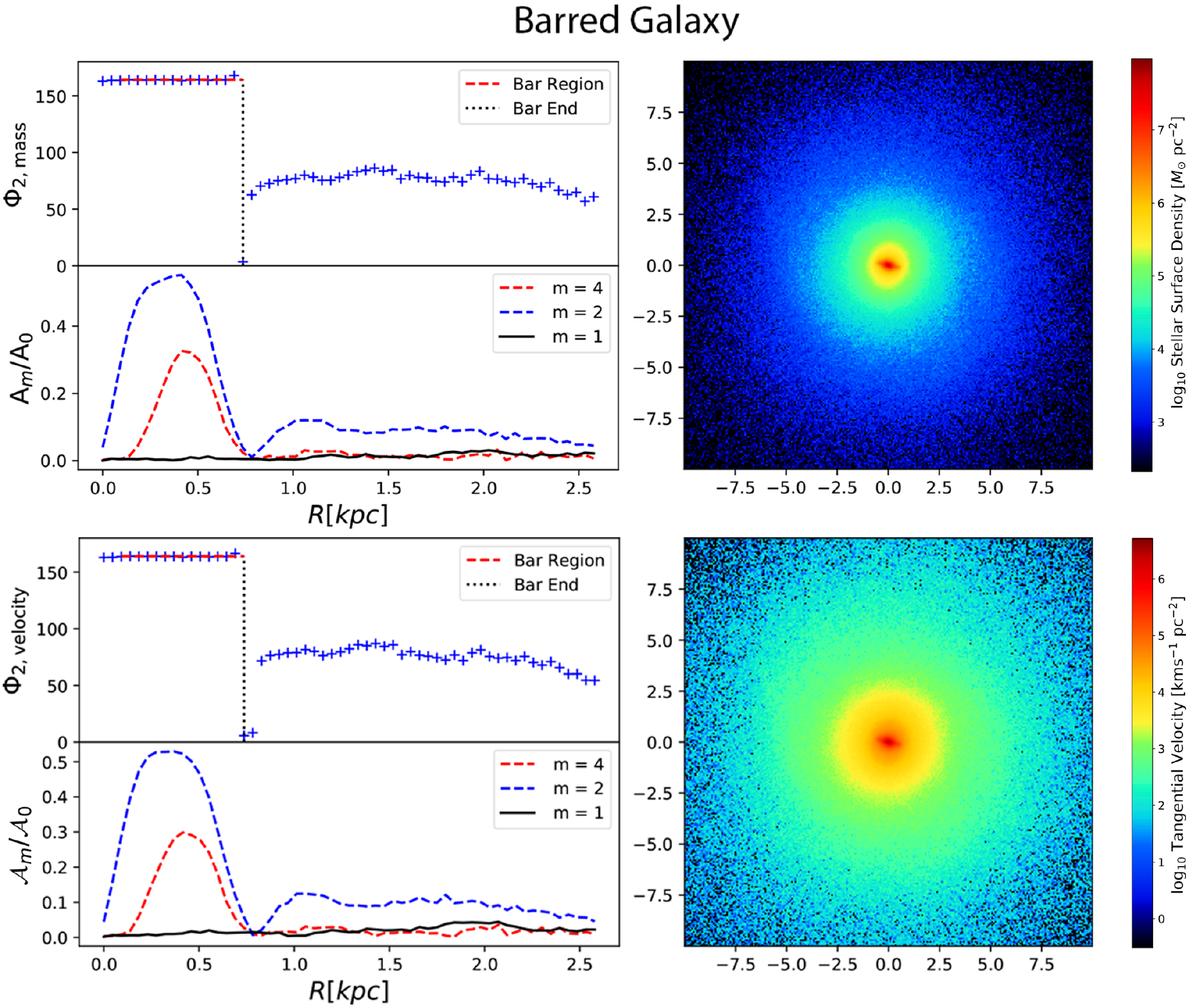}
    \caption{An example of our  procedures  used to detect a (strongly) barred \newh galaxy at \(z = 1.3\). Top row: 
    detection of a bar region by the Surface Density Harmonic Decomposition method (Section \ref{azimuthal_analysis}). On the left, we present a plot of the Fourier phase $\Phi_{2}$ weighted by stellar mass as a function of radial distance - as used to determine the bar region - along with the corresponding Fourier amplitudes. On the right, we show the stellar surface density face-on projected maps of the galaxy used for a visual inspection. Bottom row: corresponding plots for detection via the Velocity Harmonic Decomposition (Section \ref{Harmonic_Decomposition}). On the left, we present the plot of the Fourier phase $\Phi_{2}$ weighted by tangential velocity as a function of radial distance along with the corresponding Fourier amplitudes. On the right, we show a face-on projection of the tangential velocity field of the galaxy which is used by the visual inspection of the stellar surface density for the surface density method. 
    The surface density and velocity maps are plotted for $20 \times 20$ kpc$^2$ (but zoomed-in for clarity) and the colour coding scale is logarithmic.}
    \label{fig:Barred_Galaxy}
\end{figure*}

\begin{figure*}
    \centering
    \includegraphics[width=0.8\textwidth]{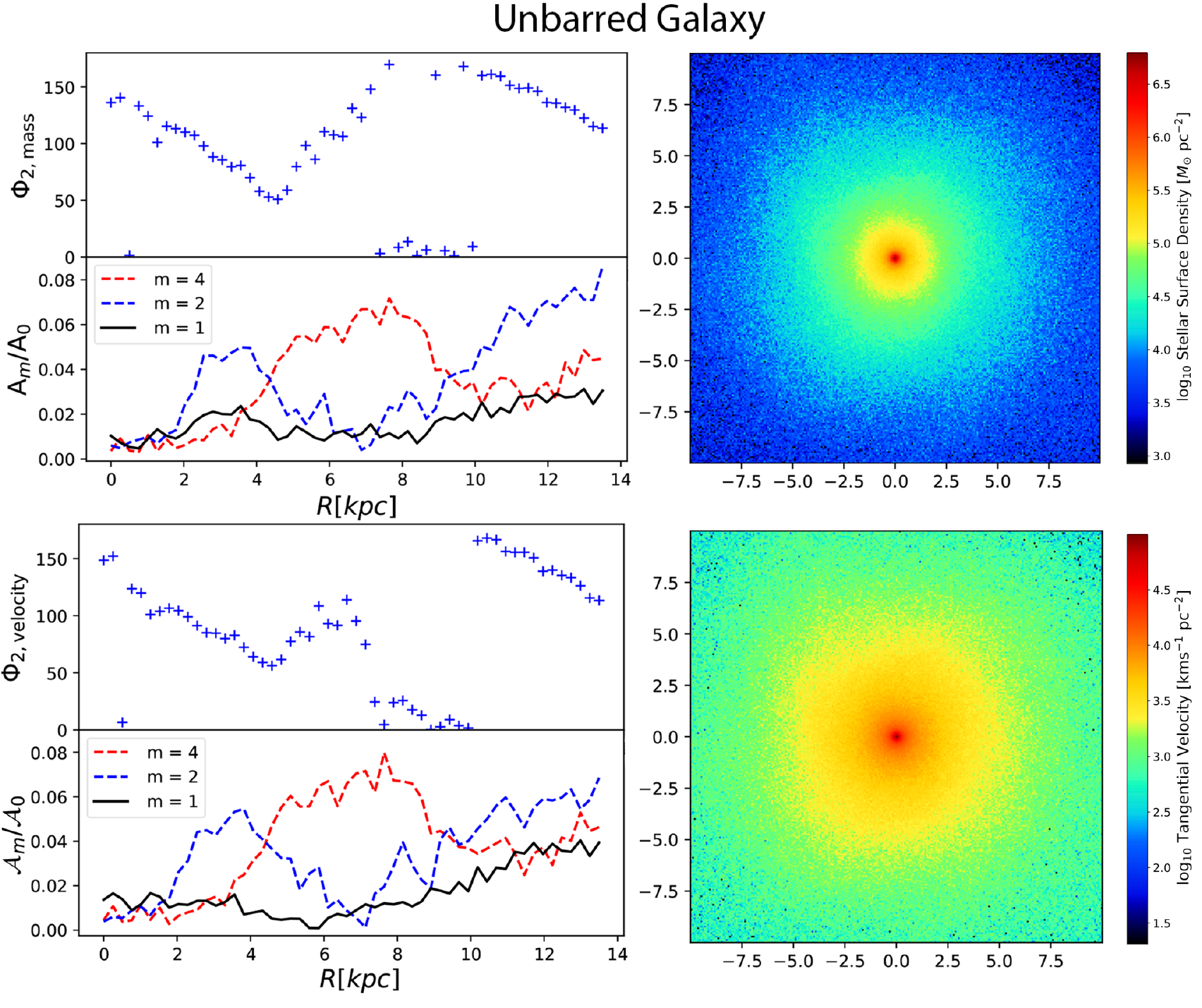}
    \caption{An example of the procedure applied to detect an unbarred \newh galaxy at z=0.25. The format is the same as in Figure~\ref{fig:Barred_Galaxy} whereby the surface density method is implemented in the top row, and the velocity method is shown in the bottom row. One can clearly see that this is an unbarred galaxy as there is no region of constant $\Phi_{2}$ either weighted by stellar mass in the surface density method, nor weighted by tangential velocity in the velocity method. As in Figure~\ref{fig:Barred_Galaxy}, the surface density and velocity maps are plotted for $20 \times 20$ kpc$^2$ (but zoomed-in for clarity) and the colour coding scale is logarithmic.}
    \label{fig:Unbarred_Galaxy}
\end{figure*}

This Fourier analysis is highly dependent on accurately finding the centre of the galaxy as when the FFT is performed, the bar search procedure starts at the origin which is supposedly the centre of mass of the galaxy. However, for cases where the origin is slightly off-centre due to substructures (more apparent in HOP selected galaxies as opposed to the AdaptaHOP which removed the majority of these), the bar search procedure will not find any bars or structure. A detailed example and insightful discussion of an off-centred stellar bars is presented in \cite{de_Swardt_2015}. To resolve for this, a shrinking sphere approach was used to effectively zoom in on the centre of the galaxy i.e. the centre of mass was first calculated within a sphere encompassing all stellar particles, the radius of this sphere was decreased by a factor of two and the centre of mass within this new sphere was calculated. The radius of the sphere was halved again and the process was repeated until the centre of mass position was consistent to a factor of $10^{-5}$ across three iterations. This process proved to visually find the centre of each galaxy and this correct centring was checked by plotting the \(m=1\) mode Fourier amplitude against radial distance. The \(m=1\) mode corresponds to lopsidedness. When centred correctly,  the \(m=1\) amplitude will be minimised at small radii\footnote{Further confirmation of our the validity of our centre-finding strategy comes from the well-defined rotation curves showing coherent rotation in all galaxies we study. See Section~\ref{subsection:rotationcurves}.}.

`Fake' bars, i.e., spheroid-dominated galaxies with moderately flattened central regions in their face-on projection \citep[see][for more details]{Kraljic_2012}, were dismissed by performing this harmonic decomposition on all three projections of the galaxy. For a given edge-on projection a disc will appear relatively flattened or `bar-like' (i.e. constant $\Phi_2(r)$). Therefore, a potential bar detected in the face-on projection had to be confirmed in both edge-on projections. However, this process proved to just be additional quality assurance on our results, as almost all spheroid-dominated galaxies were removed by the cut of \vsig described in Section~\ref{Galaxy_Selection}; on average, only two `fake' bars were removed by this process at each redshift sample.

Figure~\ref{fig:Barred_Galaxy} shows an example of a `strongly barred' \newh galaxy at \(z = 1.3\). The figure plots  the Fourier phase $\Phi_{2}$ as a function of radial distance (used to determine the bar region) along with the corresponding Fourier amplitudes. We also present the face-on projection of the stellar surface density used in the visual inspection. 
This is one of the clearest and strongest bars found in our sample - one can see the clear peaks in the \(m=2\) and \(m=4\) Fourier amplitudes withing the bar region - and we confirm this to be a barred galaxy via the detection through our velocity method (described below and with associated plots also shown in Figure \ref{fig:Barred_Galaxy}). In Figure \ref{fig:Unbarred_Galaxy}, we present the corresponding plots for an unbarred \newh galaxy at \(z = 0.25\) that is confirmed as such using the two methods. 

\subsection{Velocity Harmonic Decomposition}
\label{Harmonic_Decomposition}

 The second method for bar detection is inspired by the decompositions presented in \citet{Petersen_2019_1}. Briefly, the method uses a velocity-weighted Fourier decomposition to identify coherent velocity signatures that are consistent with barred galaxies. The stellar particles that are part of the bar \citep[the \(x_{1}\) orbit family as described in][]{Petersen_2021} dominate the velocity field, and these orbits create a detectable velocity signal that may be measured either in the average velocity field or (in the case of simulations) individual particle velocities. 

Following the surface density harmonic decomposition of Section~\ref{azimuthal_analysis}, we decompose the galaxy into Fourier components. However, we now compute the Fourier amplitude using a velocity field weighting, which we denote as $\mathcal{A}_m$. The velocity field $\mathcal{V}$ is then decomposed as
\begin{equation}
\label{Fourier_Vel_Decomposition_Eq}
    \mathcal{V} (r, \theta ) =  \mathcal{V}_{0}(r) +\sum_{m} \mathcal{A}_{m}(r)\cos (m\theta - \Phi_{m}(r)),
\end{equation}
as in equation~\eqref{Fourier_Decomposition_Eq}.

The implementation of the method to the particle data starts the same way as in the primary method: the galaxy is first rotated and centred using the  method  sketched in Section~\ref{azimuthal_analysis}. The Fourier amplitudes $\mathcal{A}_m$ are computed using the  tangential velocity field. As in the surface density decomposition, we trace the Fourier phase \(\Phi_2(r)\) in radius to find a `bar region' of constant phase. The requirements for detection are as in the surface density harmonic method (\(\Phi_2(r)\) must be constant within some radial range to \(\pm5^\circ\) tolerance).

In our application, the bar search started at a radius of 0.5 kpc and the maximum start of the search for a bar was set at 2 kpc so as to  mitigate variations caused by central structure. If the algorithm detects a region of constant Fourier phase \(\Phi_2(r)\), it is flagged to be the putative bar region -- only if the bar region covered at least 1 kpc. To ensure a particular peak in the Fourier amplitude was not noise induced, we require that the value of \(\mathcal{A}_2/\mathcal{A}_0\) at that particular radial distance remain above some threshold for significance. This threshold was set at \(\mathcal{A}_2/\mathcal{A}_0>0.02\)\footnote{To reduce the likelihood of contamination, one may lower this threshold. In our case, as we are using the velocity harmonic decomposition as a complement to the surface density harmonic decomposition, we do not take the indication of the presence of a bar as definitive proof, but rather as a candidate barred galaxy.}. 

Examples of this method used to detect both a strong barred \newh galaxy at \(z = 1.3\) and an unbarred \newh galaxy at \(z = 0.25\) are shown in Figures  \ref{fig:Barred_Galaxy} and \ref{fig:Unbarred_Galaxy} respectively. In these figures, we plot Fourier phase $\Phi_{2}$ weighted by tangential velocity as a function of radial distance - as used to determine the bar region - along with the corresponding Fourier amplitudes. We also show a face-on projection of the tangential velocity field of the galaxy which corresponds to the visual inspection of the stellar surface density for the surface density method. In Figure~\ref{fig:Barred_Galaxy}, the velocity harmonic decomposition behaves as expected: both $\mathcal{A}_2$ and $\mathcal{A}_4$ follow the expected bar signature profile, where $\mathcal{A}_4$ peaks at the end of the possible bar while $\mathcal{A}_2$ declines. In Figure~\ref{fig:Unbarred_Galaxy}, we see that the velocity harmonic profiles are not coherent in either $\mathcal{A}_2$ or $\mathcal{A}_4$, which immediately rules out the presence of a bar.

\subsection{Bar Strength}

In order to provide  a direct comparison with previous studies,
we incorporate two methods to calculate  bar strength  widely used in the literature. 
The first is that of \citet{Aguerri_1998}, that is
\begin{equation}
\label{bar_strength_eqn}
    S \equiv r_{\rm bar}^{-1} \int_{0}^{r_{\rm bar}} \frac{A_{2}}{A_{0}} dr,
\end{equation}
where A$_{0}$ and A$_{2}$ are the zeroth and second mode Fourier amplitudes respectively and $r_{\rm bar}$ is the distance at which the A$_{2}$ component becomes comparable to the level of the higher order terms of the Fourier decomposition i.e. the outer radius of the bar region. The secondary proxy for bar strength \cite[as used e.g. in][]{Diaz_2016,Rosas_Guevara_2020,Zhou_2020} is the maximum value of the ratio $A_{2}/A_{0}$ ($A_{2,\mathrm{max}}$)
within the bar region. In this paper, bar strengths are calculated using both methods, and have proven to give consistent results. Hereafter, the results presented use the integral definition (equation~\ref{bar_strength_eqn}); however, these results do not change if $A_{2,\mathrm{max}}$ is used instead. In general, care should be taken when interpreting bar strength measurements, as some theoretical studies have suggested that bar strengths derived from Fourier measurements may not accurately quantify the effect of a bar on the evolution of the host galaxy \citep[see, e.g. the comparison between apparent bar strength and the true bar potential in][]{Petersen_2019_1}.

Bars are classified depending on their bar strength, $S$, as follows;
\begin{enumerate}
    \item \textit{strong bars}, i.e. bars with a strength $S \geq 0.3$ \cite[consistent with][]{Kraljic_2012},
    \item \textit{observable bars}, i.e. bars with a strength $S \geq 0.2$. This is the typical detection limit used in observations, when studying bars at high redshift \citep{Sheth_2008}
\end{enumerate}

We note that these definitions are somewhat arbitrary as there is a continuum of bar strengths \citep{Block_2002,Whyte_2002,Menendez-Delmestre2007}. We note also that the major results of this paper do not change if we alter these values.
Any detection with strength $S < 0.2$ is neglected as at this level, it is increasingly uncertain as to whether these are bars. This is mostly due to the perturbed nature of the larger mass galaxies or the limited resolution in the lower mass regime. These ensure that any visual inspection would be challenged to identify any barred structure (see Figure~\ref{fig:barred_galaxy_grid} where barred galaxies with strength only just greater than 0.2 are already unclear).

The strength measurement is relatively well-behaved when surface density decomposition is applied. However, the meaning of $S$ is less certain in the velocity harmonic decomposition, and is likely strongly affected by noise (e.g. spatial resolution, particle number, and substructure fluctuations). 
That is, at low strength values, the false positive rate is substantial. We found that one must be particularly careful to cross-check low significance bars determined from the velocity harmonic decomposition with other methods.

\section{Redshift Evolution of  Bar Fraction}
\label{Initial_results}

The presence of a bar at increasing redshifts can indicate when galaxies have become dynamically cold and rotation dominated \citep{Sheth_2012}. Furthermore, bars are one of the most frequently and easily quantified substructures in spiral galaxies, and hence, are often used as a tracer of galaxy evolution \citep{Kraljic_2012}. A study of the redshift evolution of the bar fraction in galaxies therefore is a vital measurement when studying the evolutionary history of disc galaxies. We will now study the redshift evolution of the bar fraction in \newh. The bar analysis has been carried out at redshifts \(z = 1.3, 1.0, 0.7, 0.5\) and \(0.25\). Galaxy stellar masses range $10^{7.25} \,\msun - 10^{11.04} \,\msun$ at \(z = 1.3\) and $10^{7.28} \,\msun - 10^{11.38} \,\msun$ at \(z = 0.25\) . The errors represent
$1\sigma$ confidence intervals calculated using the (Bayesian) beta distribution quantile technique for estimating confidence intervals on binomial population proportions~\citep{Cameron_2011}.

\subsection{Bar Fractions}
\label{Bar_Fractions}

Once the morphology of galaxies is determined, the bar detection methods, as described in Section \ref{bar_analysis}, are applied to all the galaxies in the sample. If a galaxy is found to be barred, the length and strength of the bar are calculated whereby the length is simply two times the outer radius of the bar, $r_{\rm bar}$, as described above. The measurements of the bar fractions for redshifts from \(z=1.3 \) to \(z=0.25\) as calculated by both bar detection methods are shown in Table \ref{tab:Bar_Fractions}.  If we compare these values to the 30\%-70\% measured from observational \cite[e.g.][]{Eskridge_2000,Whyte_2002,Laurikainen_2004,Menendez-Delmestre2007,Marinova_Jogee_2007,Barazza_2008,Sheth_2008,Aguerri_2009,Nair_2010a,masters_2011,Masters_2012,Melvin_2014,Diaz_2016} and from simulated \cite[e.g.][]{Athanassoula_2008,Athanassoula_2012,Zhou_2020,Rosas_Guevara_2020,Rosas-Guevara_2021} studies, the disparity with our results is immediately obvious. \newh produces substantially fewer barred galaxies than is expected based on the literature. But why is there such a large disparity? We will address this in Section \ref{sec:lowbarfractions} below.

\begin{figure}
 \centering
 \includegraphics[width=\columnwidth]{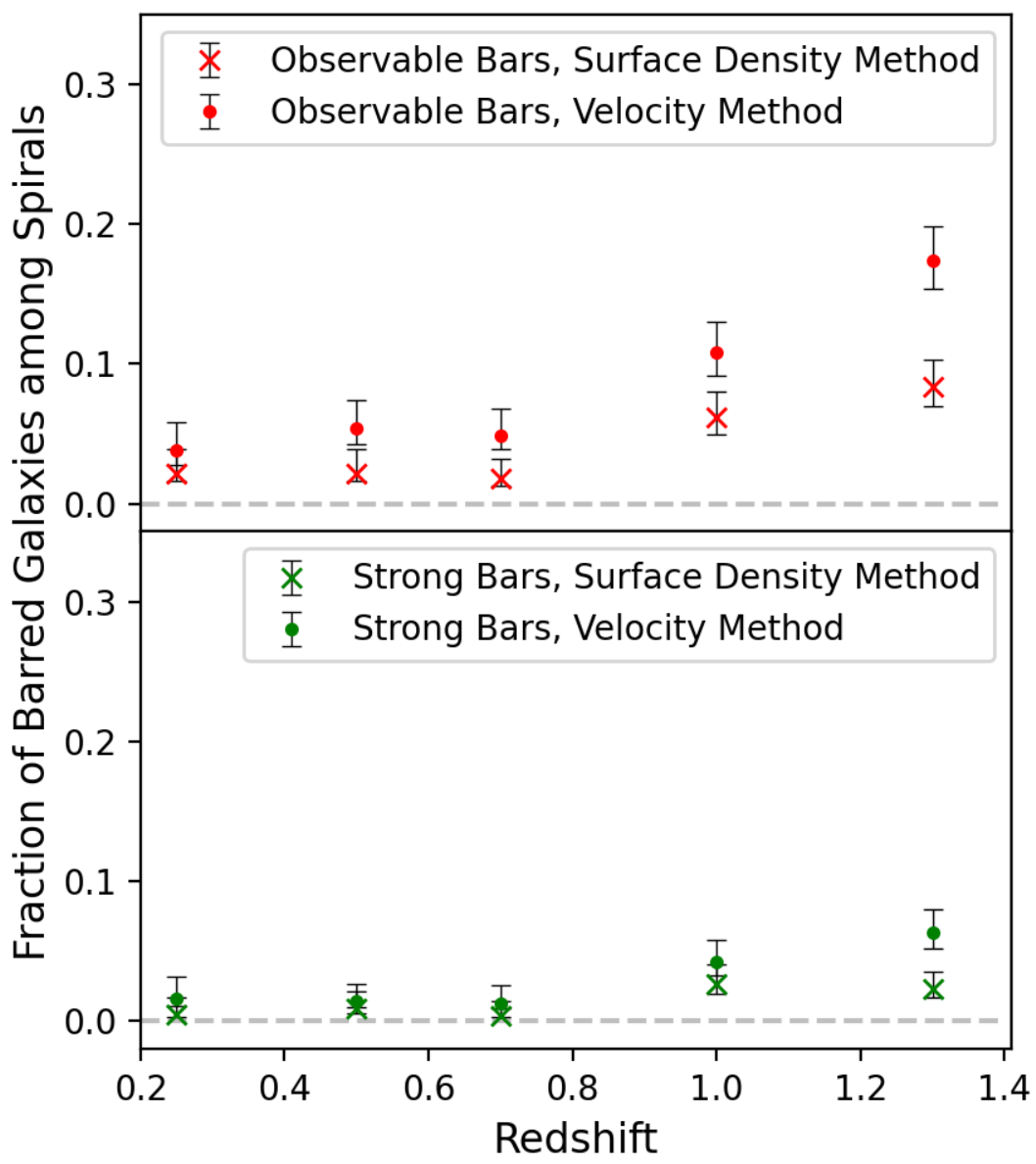}
 \caption{The redshift evolution of the bar fraction among disc-dominated galaxies across redshifts \(z = 1.3, 1.0, 0.7, 0.5\) and \(0.25\)  calculated by both  methods  described in Section \ref{bar_analysis}. The fractions from both methods for `observable bars' are shown in the top panel those for `strong bars' are shown in the bottom panel. One can see the agreement in bar fraction measurements between the two methods for `strong bars' as well as the slight discrepancies for `observable bars'.}
 \label{fig:Evolution_of_Bar_Fraction}
\end{figure}

The redshift evolution of the bar fraction is presented in Figure~\ref{fig:Evolution_of_Bar_Fraction} for both `strong' and `observable' bars. From Figure~\ref{fig:Evolution_of_Bar_Fraction}, one can see that there is little evolution in the bar fraction with redshift, with only a slight decrease in strong and observable bar fractions with decreasing redshift. One can also see that while there is good agreement in bar fractions between the two methods within the 1$\sigma$ errors, the velocity method appears to overestimate the observable bar fraction relative to the surface density method by 1-4\%. Furthermore, on a galaxy-galaxy basis there is a disparity in those galaxies that are selected as barred by the two individual methods (in cases where the galaxy would not be identified as barred by eye). These discrepancies suggest that these two methods may produce `false positives' whereby bar-like properties specific to each method are incorrectly classified as a bar. Therefore, a concurrence between the two methods is needed in order to confirm a galaxy as barred. Potential false positives will be analysed further in Sections~\ref{visual inspection results} and \ref{impact_of_substructures} by analysing our sample on a galaxy-by-galaxy basis.

\begin{table*}
\caption{The results of our analysis in terms of spiral fractions and bar fractions for `strong bars' ($S \geq 0.3$) and `observable bars' ($S \geq 0.2$) at redshifts \(z = 1.3, 1.0, 0.7, 0.5\) and \(0.25\). The bar fractions as obtained by both bar detection methods described in Section \ref{bar_analysis} are shown. The confidence intervals presented have been estimated using the (Bayesian) beta distribution quantile technique \citep{Cameron_2011}. We also present numerical values in the parenthesis.}
\centering
\begin{tabular}{ccccc}
\cline{4-5}
                      & \multicolumn{1}{l}{} &                 & \multicolumn{2}{c}{\begin{tabular}[c]{@{}c@{}}Bar Fractions\\ (Number of Bars)\end{tabular}}                                                                       \\ \hline
Redshift              & \begin{tabular}[c]{@{}c@{}}Spiral Fraction\\ (Number of Spirals)\end{tabular} & Method          & Strong Bars                                          & Observable Bars                                                                      \\ \hline
\multirow{4}{*}{1.3}  & \multirow{4}{*}{\begin{tabular}[c]{@{}c@{}}$0.570_{-0.022}^{+0.021}$\\ (299)\end{tabular}} & Surface Density & \begin{tabular}[c]{@{}c@{}}$0.023_{-0.006}^{+0.012}$\\ (7)\end{tabular}  & \begin{tabular}[c]{@{}c@{}}$0.084_{-0.014}^{+0.019}$\\ (25)\end{tabular}  \\  
                      &                       & Velocity        & \begin{tabular}[c]{@{}c@{}}$0.063_{-0.011}^{+0.017}$\\ (19)\end{tabular} & \begin{tabular}[c]{@{}c@{}}$0.174_{-0.020}^{+0.024}$\\ (52)\end{tabular}  \\ \hline
\multirow{4}{*}{1.0}  & \multirow{4}{*}{\begin{tabular}[c]{@{}c@{}}$0.543_{-0.023}^{+0.023}$\\ (260)\end{tabular}} & Surface Density & \begin{tabular}[c]{@{}c@{}}$0.027_{-0.007}^{+0.014}$\\ (7)\end{tabular}  & \begin{tabular}[c]{@{}c@{}}$0.062_{-0.012}^{+0.018}$\\ (16)\end{tabular}   \\  
                      &                       & Velocity        & \begin{tabular}[c]{@{}c@{}}$0.042_{-0.009}^{+0.016}$\\ (11)\end{tabular}  & \begin{tabular}[c]{@{}c@{}}$0.108_{-0.016}^{+0.022}$\\ (28)\end{tabular}  \\ \hline
\multirow{4}{*}{0.7}   & \multirow{4}{*}{\begin{tabular}[c]{@{}c@{}}$0.538_{-0.025}^{+0.024}$\\ (224)\end{tabular}} & Surface Density & \begin{tabular}[c]{@{}c@{}}$0.004_{-0.001}^{+0.010}$\\ (1)\end{tabular}  & \begin{tabular}[c]{@{}c@{}}$0.018_{-0.005}^{+0.014}$\\ (4)\end{tabular}   \\  
                      &                      & Velocity        & \begin{tabular}[c]{@{}c@{}}$0.013_{-0.004}^{+0.013}$\\ (3)\end{tabular}  & \begin{tabular}[c]{@{}c@{}}$0.049_{-0.010}^{+0.016}$\\ (11)\end{tabular}   \\ \hline
\multirow{4}{*}{0.5}  & \multirow{4}{*}{\begin{tabular}[c]{@{}c@{}}$0.580_{-0.026}^{+0.025}$\\ (221)\end{tabular}} & Surface Density & \begin{tabular}[c]{@{}c@{}}$0.009_{-0.003}^{+0.012}$\\ (2)\end{tabular}  & \begin{tabular}[c]{@{}c@{}}$0.027_{-0.007}^{+0.016}$\\ (6)\end{tabular}   \\  
                      &                      & Velocity        & \begin{tabular}[c]{@{}c@{}}$0.014_{-0.004}^{+0.013}$\\ (3)\end{tabular}  & \begin{tabular}[c]{@{}c@{}}$0.054_{-0.011}^{+0.020}$\\ (12)\end{tabular}   \\ \hline
\multirow{4}{*}{0.25} & \multirow{4}{*}{\begin{tabular}[c]{@{}c@{}}$0.581_{-0.028}^{+0.027}$\\ (183)\end{tabular}} & Surface Density & \begin{tabular}[c]{@{}c@{}}$0.005_{-0.002}^{+0.012}$\\ (1)\end{tabular}  & \begin{tabular}[c]{@{}c@{}}$0.022_{-0.006}^{+0.017}$\\ (4)\end{tabular}   \\  
                      &                      & Velocity        & \begin{tabular}[c]{@{}c@{}}$0.016_{-0.005}^{+0.016}$\\ (3)\end{tabular}  & \begin{tabular}[c]{@{}c@{}}$0.038_{-0.010}^{+0.020}$\\ (7)\end{tabular}   \\ \hline
\end{tabular}
\label{tab:Bar_Fractions}
\end{table*}

\subsection{Bar Properties}
\label{bar_properties}

\begin{figure}
 \centering
 \includegraphics[width=0.5\textwidth]{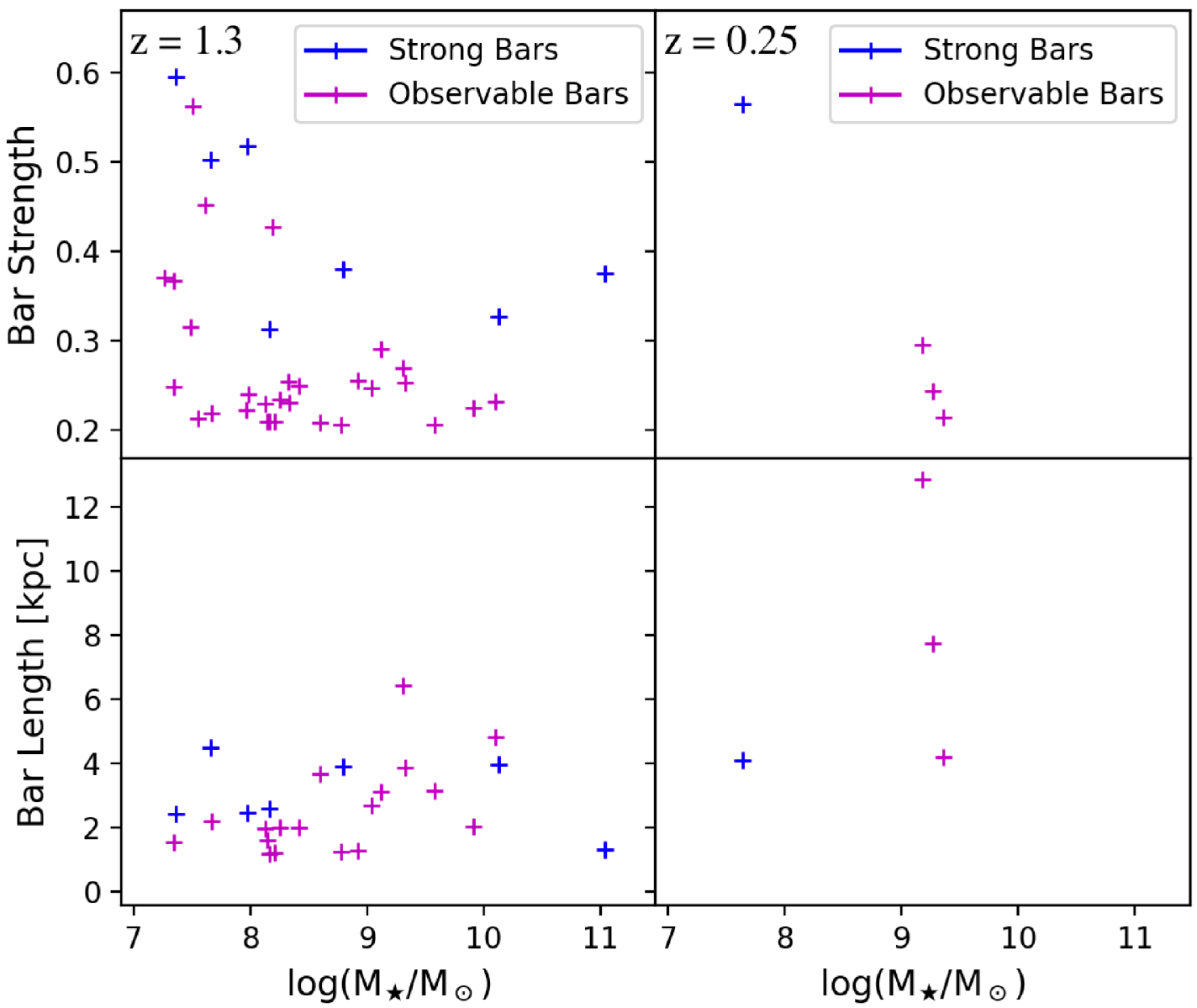}
 \caption{Figure showing bar strengths vs stellar mass (top panels) and bar lengths vs stellar mass (bottom panels) for $z$ = 1.3 (left panels) and $z$ = 0.25 (right panels). Bar strengths of both observable and strong bars are plotted for bars found in disc-dominated galaxies.  
 One can see that there is only a very weak anti-correlation between bar strengths and galaxy stellar mass and almost no correlation for bar lengths. One can also see that there is almost no evolution in these trends with redshift.
 Furthermore, these trends are seen when plotting for bars detected in all the galaxies in the sample, not just those selected as disc-dominated -- indicative of the possibility that some bias in the method causes certain galaxies to be detected as barred.
 These bars have been detected using the Surface Density Harmonic Decomposition method. A corresponding trend is found by the Velocity Harmonic Decomposition method and this plot is shown in Appendix \ref{fig:Bar_Strength_Vs_Mass_Velocity} for comparison.
 }
 \label{fig:Bar_Strength_Vs_Mass}
\end{figure}

A typical property of bars studied in the literature is that of the bar strength as a function of galaxy stellar mass. In Figure~\ref{fig:Bar_Strength_Vs_Mass} (top panels), we plot such a relationship at \(z = 1.3\) and 0.25 for bars detected using the Surface Density Harmonic Decomposition method. At \(z = 0.25\) we find a weak but statistically significant anti-correlation\footnote{Spearman coefficient of -0.335 with p-value=0.011}, whereby more strongly barred galaxies tend to be in the lower mass regime while the higher mass galaxies tend to hold weaker bars. There is very little redshift evolution of this relation with a slight decrease in the number of detected bars. Similar trends are found if we perform our analysis on all the galaxies in the sample, not just those selected as disc-dominated. And finally, comparable trends are found when using the Velocity Harmonic Decomposition method to detect bars, as shown in Figure~\ref{fig:Bar_Strength_Vs_Mass_Velocity}. These points all seem to be indicative of i) there being very few strongly barred massive galaxies in \newh and ii) there being some bias in the methods causing galaxies (that are not barred upon visual inspection - see Section \ref{visual inspection results}) to be detected as barred.

In Figure \ref{fig:Bar_Strength_Vs_Mass} (bottom panels), we also plot the length of the detected bars against stellar mass. However we find that there is no clear dependence with mass for detected bars. This is in immediate disagreement with observations; \cite{Erwin_2018,Erwin_2019} in the local universe, \cite{Barazza_2009} at intermediate redshift and \cite{Kim_2021} out to $z \sim$ 0.84 all show a clear positive correlation between bar length and galaxy stellar mass. Again, we see little to no evolution in this relation with redshift, this time in agreement with \cite{Kim_2021}.
 
Our initial bar search criteria required a potential bar region to be at least 1 kpc in length before classifying the galaxy as barred. However, due to the spatial resolution of the \newh simulation, we should expect to be able to resolve bars shorter than this. Despite this, by reducing this criterion on the bar length, we detect only one extra short bar (that having 0.5 kpc \(<\) bar length \(<\) 1 kpc) compared to longer bars (those with lengths \(>\) 1 kpc). This occurs when using the velocity method and is plotted in Figure \ref{fig:Bar_Strength_Vs_Mass_Velocity} for completeness, demonstrating that our results do not change when allowing also for short bars. 

In order to decipher the causes behind any biases in our bar detection methods, we will inspect galaxies visually as well as comparing detection methods on a galaxy-galaxy basis in Section \ref{sec:technicalities}. We will also compare our results in more detail to those of the literature in Section \ref{Discussion} discussing their relevance and the lessons for future research. Let us now address the large disparity between our measured bar fractions and those of the literature - why the low bar fractions in \newh and is there a missing bar problem in cosmological simulations?

\section{A missing bar problem?}
\label{sec:lowbarfractions}

While the biases that we will discuss in Section \ref{sec:technicalities} may explain some discrepancies between  measured bar fraction in observations and simulations, in particular at high redshift and/or at low stellar masses, the significant lack of bars in the \newh simulation seems to point to a more fundamental problem. 

Let us first discuss resolution effects in Section~\ref{subsec:resolution}. We then present simple dynamical analysis of the disc galaxies at \(z=1.3\) and \(z=0.25\).  For each of the disc galaxies, we extract the gas and dark matter components within 6\(R_{\rm eff}\). For galaxies above \(\log(M_{\star}/\msun)>10.2\), we extract the dark matter component out to 30\(R_{\rm eff}\) to account for the extended nature of these galaxies.  We  construct rotation curves, defining summary statistics (Section~\ref{subsection:rotationcurves}), and then analyse the sample of disc galaxies, making simple dynamical arguments for the low bar fraction (Section~\ref{subsection:sample}). We  compute linear growth rates for  composite models matching the population to quantify the dynamical state of the discs
(Section~\ref{subsec:linearstability}).

\subsection{Resolving secular processes}
\label{subsec:resolution}

The resolution required to properly treat secular processes is not straightforward. In general, one may consider the role of both mass and spatial `resolution' in cosmological simulations. Both are critical for correctly modelling secular processes, which can strongly be amplified by the disc's temperature, should the disc's thickness 
be resolved \citep{Fouvry2015}. 

The mass resolution, or the number of particles, has been identified as a crucial ingredient  for the accurate modelling of complex secular processes in galaxies such as bar formation, spiral arms or pseudo-bulges. For instance, by studying the details of resonant interactions, \cite{WeinbergI_2007} derived explicit criteria for the number of particles required to capture accurately these dynamical processes in N-body simulations. Typically, for a Milky Way-like galaxy, $10^7-10^8$ particles are required  to minimise the fluctuations in the gravitational potential  so as  to sufficiently populate   regions near  resonances.  This should prove in particular  relevant to  model dynamical friction from the dark halo onto the  bar. Yet modern simulations, and in particular zoom-in simulations, typically employ a number of particles that approaches this requirement \citep[even though  they might not correctly model the full disc dynamics; see][for more discussion]{WeinbergII_2007}. 

However, in the context of bar formation in cosmological simulations, spatial resolution appears to play a critical role. In a meta-analysis of the literature, the bar fraction in zoom-in simulations of Milky Way-like galaxies that have spatial resolution  better than $\sim$100 pc is essentially zero. Some recent examples include, e.g. \textsc{Vintergatan} \citep{Agertz_2020} with a resolution of $\sim$ 20 pc, \textsc{Fire-2} \citep{Hopkins_2018}, simulations of \cite{Nunez_2021}, or Galactica \citep{Park_2021} with a resolution comparable to that of \newh.

Zoom-in simulations that manage to form bars typically choose lower spatial resolution, either motivated by `optimal' softening length arguments \citep{Power_2003} or owing to mass resolution. For example, \cite{Zana_2018} uses the softening length of 120 pc, the Auriga project \citep{Grand_2017,BlazquezCalero_2020} adopts the physical softening length for stars of 369 pc and for some older zoom-in simulations such as e.g. \cite{Scannapieco_Athanassoula_2012}, it is 1.4 kpc for gas, stars and DM components, in \cite{Agertz_2011}, the maximum level of refinement reaches a physical resolution of 170 pc, and the simulations used in \cite{Kraljic_2012} have the spatial resolution of 150 pc.

As will be discussed in Section~\ref{visual inspection results}, the most prominent barred galaxy at \(z=1.3\) belongs to the most massive disc at that redshift\footnote{Given that the mass of stellar particles are the same, the galaxy therefore has the largest number of stellar particles -- apparently near the regime \citealt{WeinbergI_2007} would suggest for resolving some secular processes.}. Is this a coincidence? 
Given that this galaxy is found to be barred at \(z=1.3\), 1.0 and 0.7, it is highly unlikely that this detection is a statistical fluke. Thus,  \newh seems  able to produce the conditions for bar formation under some circumstances. However, its presence  cannot be fully explained by mass resolution: a more massive (and therefore equally high mass-resolution) disc is definitively unbarred  at \(z=0.25\).

Finally, the (star formation recipe and resolution-dependent) number of massive clumps within the disc may impact the secular resilience of bars. This issue can be qualitatively addressed via the  so-called  Hamiltonian mean field model \citep{Pichon1993,Antoni1995}.  This  involves computing orbital diffusion coefficients within the separatrix defined by the bar, using the Balescu-Lenard quasi-linear theory \citep{Heyvaerts2010,Fouvry2018}, which captures  the impact of the number of  orbit-averaged clumps, given their relative contribution to the bar's  self-gravity. \cite{Benetti2017} compute these coefficients  as a function of bar strength,  and find   (their figure~6) that i) below the separatrix, diffusion is fairly efficient for weak bar, while it is much less so for strong bars; ii) at the separatrix, the diffusion coefficient vanishes; iii) in the strong bar limit, wakes weaken diffusion (i.e. their Balescu-Lenard predictions versus  Landau's).  While their setting is idealised, with a slightly different geometry, their results seem consistent with the naive expectation that a weaker bar will dissolve more easily  through resonant encounters, whereas its strong  counterpart will more easily sustain massive clump formation. It remains to be seen that i) the lifespan of these clumps is long enough for  relaxation processes to operate, ii)  the strength of resonant encounters (quantified by these diffusion coefficients) are sufficient for  clumps to jump the separatrix and iii) the fraction of the bar's mass they drag is enough to eventually dissolve the bar. More prosaically, the number and size of stellar clumps may vary with increased resolution or change in star formation modelling, which would artificially impact  bar resilience.

In closing, whether the spatial resolution and/or the choice of softening length and subgrid physics in simulations plays a key role in the presence or absence of bars should be the topic of further investigations. In the next sections, we examine two summary  statistics and connect them with known secular processes responsible for bar formation in order to probe alternative, physically motivated reasons for the lack of bars in \newh.

\subsection{Insights from rotation curves}\label{subsection:rotationcurves}

One of the most straightforward ways to analyse the dynamics of a galaxy is through the rotation curve, or equivalently, the circular velocity curve. We construct the circular velocity curve from the mass enclosed (\(v_c(r)=\sqrt{G{\rm M}(<r)/r})\) profile of the stellar, gas, and dark matter components, both separately and combined. While imperfect\footnote{In particular, the assumption of a spherical distribution for the stellar and gas components introduces a small (up to $10\%$) bias in derived quantities.}, we find through visual inspection and comparison of the circular velocity to the measured mean tangential velocity of the disc stars that the circular velocity curves constructed through these naive means is satisfactory for dynamical inference. We will hereafter drop the `\(c\)' subscript and refer to the circular velocity simply as `\(v\)'.

\begin{figure}
    \centering
    \includegraphics[width=3.25in]{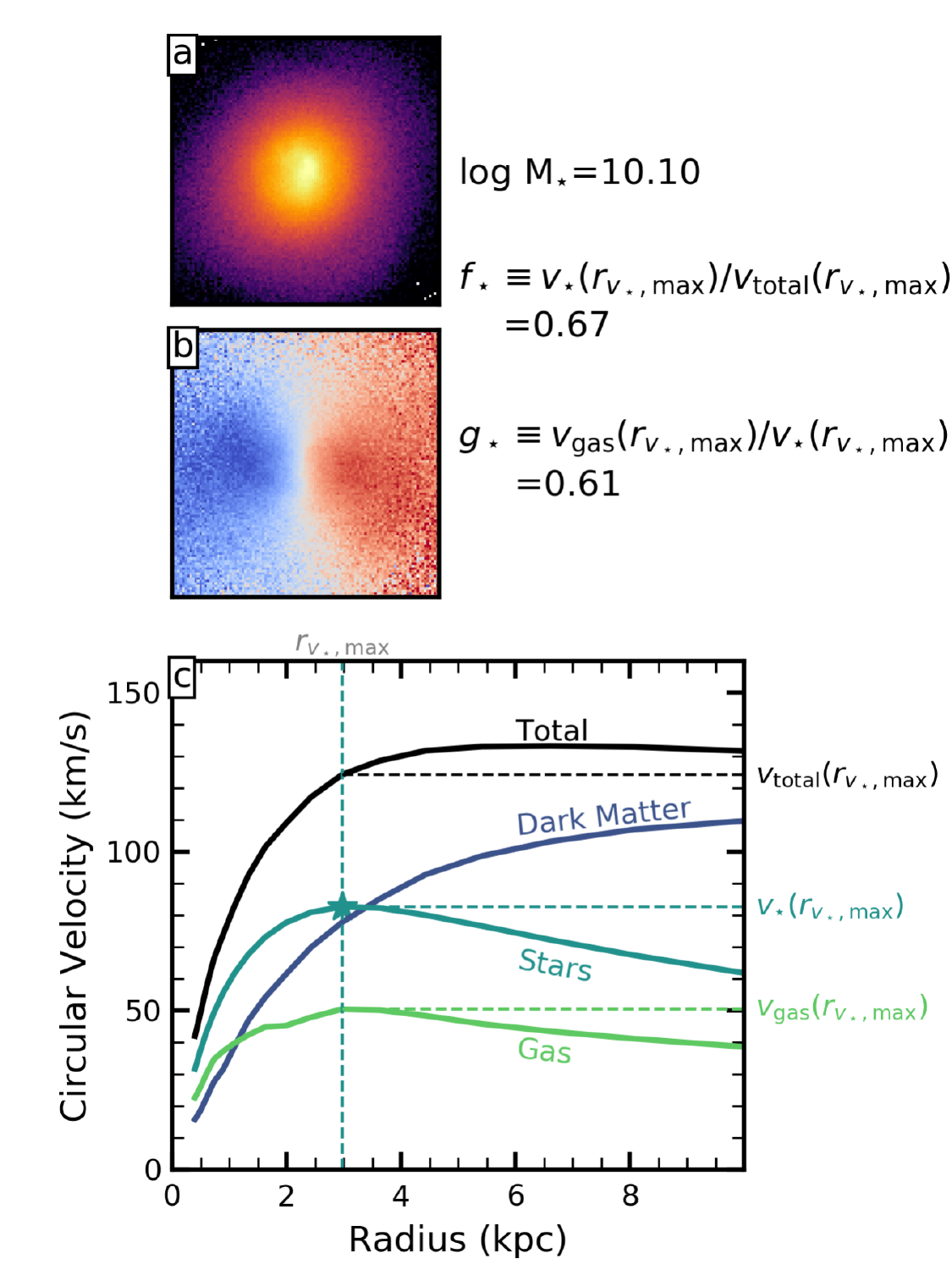}
    \caption{Face-on stellar surface density, velocity field, and rotation curve for an example galaxy at \(z=0.25\). The surface density resembles an exponential disc with a small bulge component. We also show the velocity field, which exhibits regular rotation. Panels a and b are 20 kpc per side. Panel c is the circular velocity curve, with listed and derived quantities as defined in the text. We show the total circular velocity curve in black, as well as the contributions from the dark matter, stellar, and gas components in coloured lines as marked. The location of the peak circular velocity contribution from the stellar disc is marked with a star.
    \label{fig:pedagogical_rotcurve}}
\end{figure}

Figure~\ref{fig:pedagogical_rotcurve} is a pedagogical demonstration of summary statistics we derive for each galaxy from the circular velocity curve. We define three key quantities in addition to the total stellar mass:
\begin{enumerate}
\item The radius at which the stellar contribution to the total circular velocity is maximised, \(r_{v_{\star,{\rm max}}}\).
\item The contribution to the total rotation curve by the stellar component at \(r_{v_{\star,{\rm max}}}\), which gives a measure of how `maximal' the stellar component of a galaxy is, \(f_\star\equiv v_{\star}~(r_{v_{\star,{\rm max}}})/v_{\rm total}~(r_{v_{\star,{\rm max}}})\).
\item The ratio of the gas rotation curve value to the stellar rotation curve value at \(r_{v_{\star,{\rm max}}}\), which gives a measure of the gas content in the galaxy relative to the stellar content, \({g_\star\equiv v_{{\rm gas}}~(r_{v_{\star,{\rm max}}})/v_{\star}~(r_{v_{\star,{\rm max}}})}\). 
\end{enumerate}
In a pure exponential disc, \(r_{v_{\star,{\rm max}}}\) corresponds to \(2.2R_d\), where \(R_d\) is the disc scale length. This is independent of the mass in the exponential disc, ${\rm M_{disc}}$. The addition of a spherical bulge with mass ${\rm M_{bulge}}$ can alter the value of \(r_{v_{\star,{\rm max}}}\): at fixed \(R_d\), as the bulge fraction \(f_{\rm bulge}\equiv {\rm M_{bulge}}/({\rm M_{bulge}}+{\rm M_{disc}})\) is increased, \(r_{v_{\star,{\rm max}}}\) decreases\footnote{Changing the geometry of the bulge, such as decreasing \(\alpha\) in a  \(\rho_{\rm bulge}\propto r^{\alpha}\) bulge can also affect the location of \(r_{v_{\star,{\rm max}}}\) at fixed \(f_{\rm bulge}\), but in reasonable toy models we find this to be a second-order effect when compared to changing \(f_{\rm bulge}\).}. In the limit where \({f_{\rm bulge}\to}1\), \({r_{v_{\star,{\rm max}}}\to}0\). In practice,  \(r_{v_{\star,{\rm max}}}=0\) (where the bulge dominates the rotation curve of the galaxy) is achieved at \(f_{\rm bulge}\gtrapprox0.35\). When \({r_{v_{\star,{\rm max}}}\to}0\), \(f_\star\) increases dramatically towards unity.

\begin{figure*}
    \centering
    \includegraphics[width=6.7in]{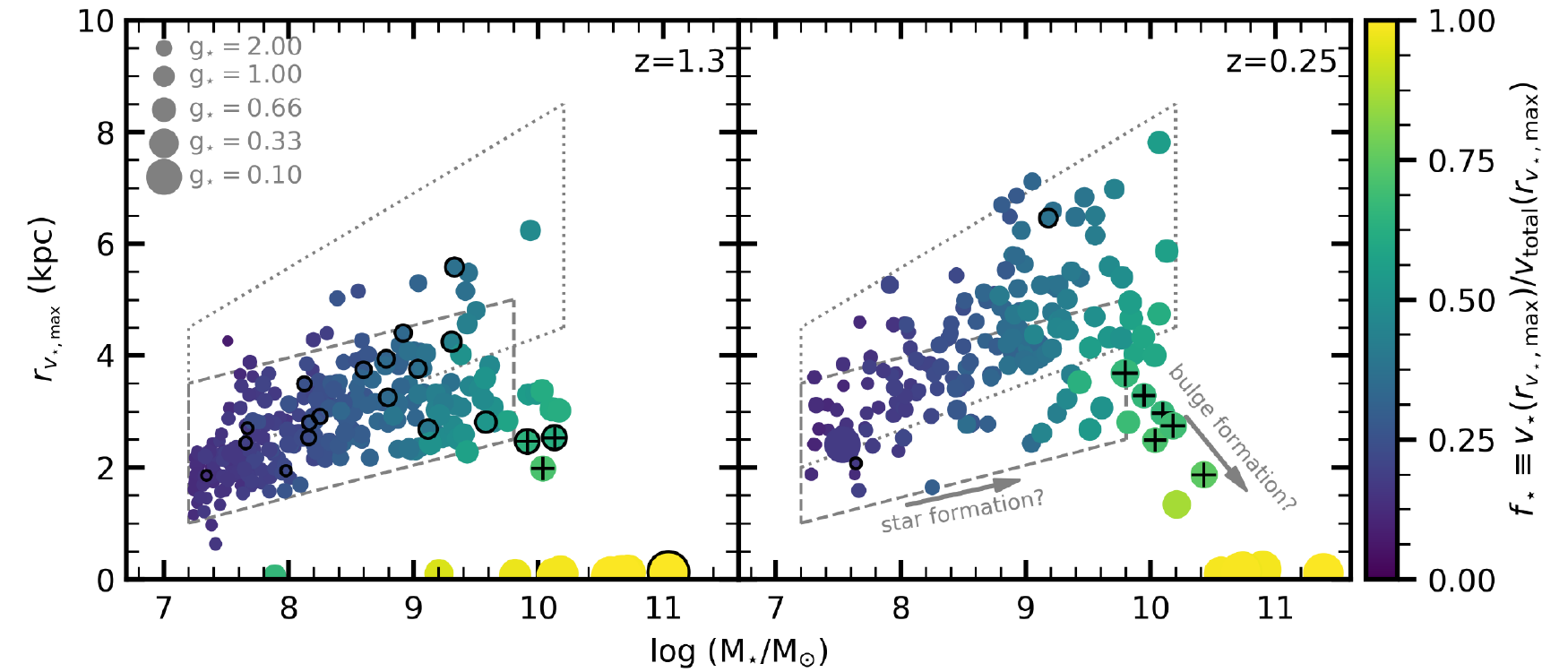}
    \caption{Two snapshots at different redshifts of disc galaxies in the $\log\left({\rm M}_\star/\msun\right)-r_{v_{\star,{\rm max}}}$ plane. Colours correspond to \(f_\star\); symbol sizes correspond to \(g_\star\) (all quantities are defined in text). Galaxies identified as barred are outlined in black. Galaxies that fall in a region of `possible secular bar formation' are indicated with a black `+' (\(0.4<f_\star<0.8\) and \(g_\star<0.66\)). Most galaxies are excluded from the region of possible bar formation owing to either being relatively low stellar mass or by being bulge dominated. The left panel shows the highest redshift considered in this study, \(z=1.3\). The right panel shows the lowest redshift considered in this study, \(z=0.25\). Two possible evolutionary tracks are shown: first, star formation is expected to be responsible for the increase in $r_{v_{\star,{\rm max}}}$ with stellar mass. Second, bulge formation may be responsible for the dearth of galaxies with large $r_{v_{\star,{\rm max}}}$ at higher stellar masses. To guide the eye and easily follow the redshift evolution in this parameter space, in each panel we draw rough polygons around the bulk of the galaxies (dashed outline polygon corresponds to \(z=1.3\), dotted outline polygon corresponds to \(z=0.25\)). \label{fig:summarystatistics}}
\end{figure*}

For \(f_\star\) and \(g_\star\), we can draw insights from observational and theoretical literature for typical values that may promote secular bar formation. Local disc galaxies, including some hosting bars, exhibit \(f_{\rm baryonic}\equiv f_{\star}\sqrt{1+g_{\star}^2} \in (0.4,0.7)\), with \(\langle f_{\rm baryonic}\rangle=0.57\) \citep{Martinsson_2013}. As the majority of the galaxies in the \citet{Martinsson_2013} sample have relatively low gas content\footnote{The maximum gas-to-stellar fraction (including both atomic and molecular gas) in this sample is 40\%, which corresponds to \(g_\star=0.63\).}, a fair comparison for benchmarking to our simulations is \(f_\star \approx f_{\rm baryonic}\). \citet{Athanassoula_2013} found that increasing the gas content in isolated disc galaxies acted to slow the bar formation process, such that gas-dominated galaxies were near-axisymmetric for longer periods of time. Both linear stability studies \citep[e.g.][]{Pichon_1997} and isolated simulations (Petersen et al. in prep) of disc galaxies indicate that at low \(f_\star\), bar formation is prohibited by the dominance of the dark matter halo. Conversely, when \(f_\star\) is near unity owing to the presence of a large bulge driving the radius of the maximum of the stellar velocity curve to the centre of the galaxy, the galaxy dynamics may be stabilised against non-axisymmetric structure formation \citep{Ambastha_1982}. Taken together, we define a broad literature-derived space for `possible secular bar formation' as \(0.4<f_\star<0.8\) and \(g_\star<0.66\). 

\subsection{Insights from the entire sample} \label{subsection:sample}

Combining the summary statistics produces a four-dimensional space with the power to produce dynamical insights. In Figure~\ref{fig:summarystatistics}, we show disc galaxies in the \(\log\left(\mstar/\msun\right)-r_{v_{\star,{\rm max}}}\) plane at the highest redshift we consider (\(z=1.3\), left panel) and the lowest redshift we consider (\(z=0.25\), right panel). As in \citet{Jackson_2021}, we find that the dark matter content of the disc galaxies is in agreement with observations.

Through a linear regression analysis, we find a positive correlation between \(r_{v_{\star,{\rm max}}}\) (or \(R_d\)) and \(\log\left({\rm M}_\star/{\rm M}_\odot\right)\) for galaxies with \(\log\left({\rm M}_\star/{\rm M}_\odot\right)<9.5\) at both redshifts\footnote{This partitioning of the sample is also degenerate with cuts on either $f_\star$ and $g_\star$.}. We interpret this relationship as a natural consequence of the galaxy formation process, where the scale length of the galaxy increases as the stellar mass of the galaxy increases through star formation. Indeed, a comparison of the two redshift snapshots suggests that galaxies generally increase \(r_{v_{\star,{\rm max}}}\) (and thus scale length) while increasing mass. We also see generally increasing \(f_\star\) and decreasing \(g_\star\) from \(7<\log\left({\rm M}_\star/{\rm M}_\odot\right)<9.5\).

However, above \(\log\left({\rm M}_\star/{\rm M}_\odot\right)=9.5\), a linear regression analysis finds a slight {\sl negative} correlation between \(r_{v_{\star,{\rm max}}}\) and \(\log\left({\rm M}_\star/{\rm M}_\odot\right)\) in the $z=0.25$ sample, and no correlation in the $z=1.3$ sample. Visual inspection of rotation curves indicates that this is a result of increasing \(f_{\rm bulge}\), towards the limit where the galaxy rotation curve peaks at small radii as a result of the presence of a bulge (the lower right corner of Figure~\ref{fig:summarystatistics}). The apparent qualitative break from the positive trend seen at low mass ($\log\left({\rm M}_\star/{\rm M}_\odot\right)\leq 9.5$) suggests that galaxies in \newh may be in the process of forming large bulges above a certain mass threshold\footnote{Complicating our study and the relationship to previous works, the measurement of bulges in models is far from settled. Both kinematic \protect{\citep[e.g.][]{Park_2021}} and photometric \protect{\citep[e.g.][]{BlazquezCalero_2020}} methods are regularly employed \protect{\citep[and sometimes both, e.g.][]{Gargiulo_2019}}. While we have not attempted a detailed decomposition, the rotation curve in our model galaxies indicate significant central mass concentrations that set \(r_{v_{\star,{\rm max}}}\), which we refer to as `large bulges'.}. Evidently, some bulge formation proceeds at high redshift: several galaxies already exhibit prominent bulges by \(z=1.3\). We will study the formation and evolution of bulges in an upcoming work.

We mark the points of galaxies in Figure~\ref{fig:summarystatistics} that fall in the `possible secular bar formation' regime defined above with black `+' symbols. Of the sample of 128 (200) disc galaxies at \(z=0.25\) (\(z=1.3\)), we find only {\sl six} (three) that satisfy these criteria. Relaxing the \(g_\star\) threshold to \(g_\star<1.0\) increases the number in the `candidate' bar space dramatically (but does not capture any additional galaxies classified as barred in the sample). Widening the \(f_\star\) range to extend down to \(f_\star>0.3\) does not add any additional galaxies in the candidate bar space if \(g_\star\) is held fixed. Visual inspection of the six \(z=0.25\) galaxies classifed as barred (including the galaxy in Figure~\ref{fig:pedagogical_rotcurve}) confirms that these galaxies are not barred, and prompts the question of whether the defined thresholds are too conservative: one explanation is that the actual bar-forming region in a theoretical \(f_\star-g_\star\) plane may be tighter than we conjecture above. One could also argue that \(r_{v_{\star,{\rm max}}}\) as measured for disk galaxies in \newh are unrealistic. However, this does not seem to be a plausible explanation as the effective radii (and therefore also the maximal radii) of \newh galaxies are in a broad agreement with observations \citep[see][]{Dubois_2020}.

We mark the galaxies with detected bars (the putative bars as confirmed via a concurrence between our methods - see Section \ref{visual inspection results}) in Figure~\ref{fig:summarystatistics} with black outlines. This results in 21 galaxies at \(z=1.3\) and 2 galaxies at \(z=0.25\). The galaxies do not appear to preferentially reside in any region of the \(\log\left({\rm M}_\star/{\rm M}_\odot\right)-r_{v_{\star,{\rm max}}}\) plane. None of the galaxies detected as barred at \(z=1.3\) are detected as barred at \(z=0.25\). The formerly barred galaxies do not appear to have obviously undergone any `special' evolution, or reside in a unique location in \(f_\star-g_\star\) space.

In particular, the defined `possible secular bar formation' region in parameter space does not appear to hold any particular importance for \newh galaxies. As the region is simply a space where isolated galaxy studies have previously identified bars, it does not preclude bar formation in regions outside of this space. The identification of bars in \newh outside the possible secular bar formation region identified in isolated studies suggests that the bar formation in \newh might be driven by different mechanisms than in isolated studies. We have little information about possible regions for bars triggered by interactions with other galaxies. Such a study is beyond the scope of this work, but cosmological simulations have a unique opportunity to explore bar formation beyond the secular mode traditionally studied in bar dynamics. We conclude that the bars identified in \newh do not resemble secularly formed bars as identified in previous studies of isolated galaxies.

\subsection{Insights from linear-response growth rates}
\label{subsec:linearstability}

Let us now present the trends from a linear response toy model tailored to provide insight into the lack of bars in \newh. 
We follow \cite{Aoki1979}, which studied the stability of razor thin, gaseous, isolated discs, while including varying bulge and halo components \citep[see also][]{Ambastha_1982}. The construction of the model is detailed in Appendix~\ref{sec:Toymodel}. The model is parameterised by four ratios: ${p=M_{\rm bulge}/(M_{\rm disc}+M_{\rm bulge})}$, ${q=M_{\rm disc}/(M_{\rm disc}+M_{\rm halo})}$, \(a_\mathrm{b}/ a_\mathrm{d}\), and \(a_\mathrm{h}/ a_\mathrm{d}\), where \(a_{\{\mathrm{b},\mathrm{d},\mathrm{h}\}}\) are the scale lengths of the bulge, disc, and halo respectively and $M_{\rm halo}$ is the mass of the halo corresponding to the virial mass. With this parameterisation, \(a_\mathrm{d}\) and \(M_{\rm disc}\) may be readily scaled out. While galaxies in this model live in a four-dimensional space, we find it is reasonable for \newh galaxies to reduce the four-dimensional space to a two-dimensional space parameterised by \((p,q)\) by using typical values for \(a_\mathrm{b}/ a_\mathrm{d}\) and \(a_\mathrm{h}/ a_\mathrm{d}\) motivated by a mapping from the \newh galaxies to the model (see Appendix~\ref{sec:mapping}).

\begin{figure}
    \centering
    \includegraphics[width=\columnwidth]{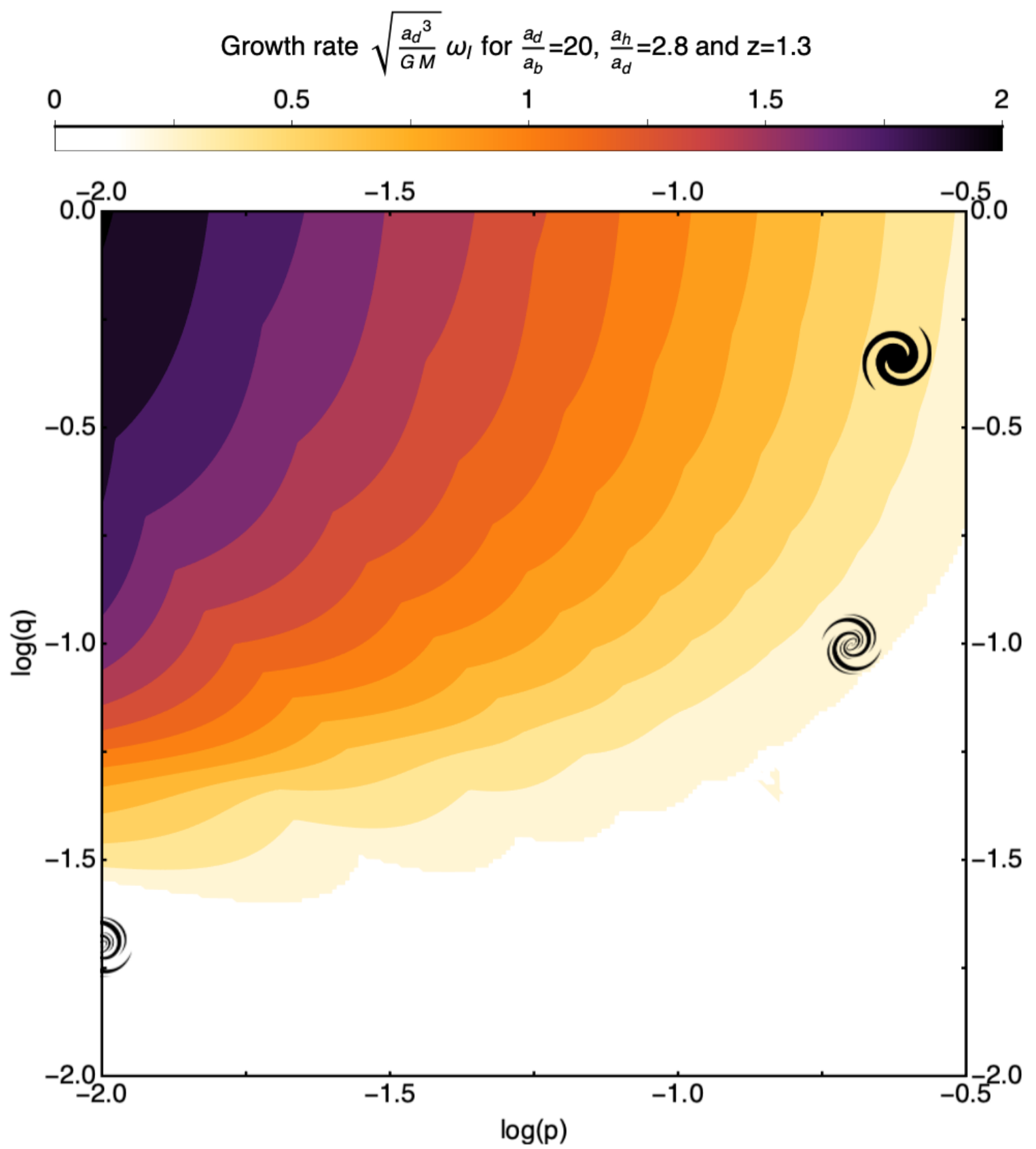}
    \caption{
    The fastest growth rate of ${m=2}$ modes in the Kuzmin-Toomre-Plummer fiducial model   as a function of the two parameters, $p$, the bulge fraction, and  $q$, the disc fraction for the quoted values of scale lengths. The bulge, disc and halo scale lengths are chosen to match those of \newh. The completely white region corresponds to (ad hoc) values of $\omega_{\mathrm{I}}$ less than ${0.1\sqrt{{G M\!/\ad^3}}}$, which would lead to too slow bar mode growth. As a function of cosmic time, via accretion, star formation and mergers, the galaxies  move within this diagram and may cross this threshold, triggering spontaneous bar formation. We place three representative points within the space:
   the bottom left represents  a typical barless bulgeless galaxy, 
   the central one a typical barless bulge-hosting galaxy, and
  the top one showing the approximate location of the most prominent barred galaxy in the sample at \(z=1.3\). The \newh does not produce bi-symmetric unstable discs, with the exception of that one galaxy, which 
  only hosts a bar for a fraction of a Hubble time.
  }
    \label{fig:disc-stab}
\end{figure}

We focus on  linear instabilities possibly leading to bar formation: the bi-symmetric  $m=2$ mode. Figure~\ref{fig:disc-stab} plots the fastest  growth rate,  $\omega_{\mathrm{I}}$, of a sequence of models parametrised by $p$ and $q$, with \(a_\mathrm{b}/ a_\mathrm{d}=20\) and \(a_\mathrm{h}/ a_\mathrm{d}=2.8\). Light (resp. dark) contours corresponds to slow (resp. fast) growth.
The inverse of this growth rate quantifies the time it takes for a 
linear instability to grow (in units of the dynamical time).
As seen on this figure,  DM-dominated and bulge-hosting galaxies both have slow growth rates (lower right light region)\footnote{We set up a growth time threshold  at a fraction, $\eta$ of the dynamical time ${ \sqrt{\ad^3/G\! M_\star}}$ which,  we somewhat arbitrarily set at $\eta =10$.  For a typical stellar mass of $10^{10} \, \msun$ and $\ad=3$ kpc, this corresponds to 0.25 {Gyr}.}, unlikely to secularly grow strong non-linear bars, in qualitative agreement with the results in Section~\ref{subsection:sample}.

With the mapping derived in Appendix~\ref{sec:mapping}, we find that the bulge-less and bulge-hosting galaxies may be roughly split into distinct populations in \((p,q)\) space. We therefore select `typical' values to represent the two populations. We place an open galaxy symbol on Figure~\ref{fig:disc-stab} at \((\log p,\log q)=(-2,-1.7)\) representing the typical bulge-less galaxy in the sample at \(z=1.3\). We also place an open galaxy symbol at the location of the typical bulge-hosting galaxy  at 
\((\log p,\log q)=(-0.7,-1.0)\). 
Comparing the bulge-less and bulge-hosting galaxies, we find that while the galaxies would be considered unstable with the modestly larger \(\log q\) values than the bulge-less counterparts, the bulge acts as a stabiliser when interpreted in the framework of this toy model. Interestingly, the most prominent barred galaxy at \(z=1.3\) is by far the most massive disc in the \(z=1.3\) sample, and lies at 
\((\log p,\log q)=(-0.62,-0.34)\), 
in a region where the bulge mass is too small for the bulge to stabilise the disc against bar formation. This galaxy is shown in Figure~\ref{fig:Barred_Galaxy} and is shown  as a  filled galaxy symbol in Figure~\ref{fig:disc-stab}. However, tracking the same galaxy forward to lower redshift reveals that \(\log q\) decreases while \(\log p\) modestly increases, both of which would act to stabilise the galaxy against bar formation, and indeed, the observed bar is gone by \(z=0.7\). 

At the level of this admittedly crude model, we show that nearly all \newh discs fall in the linearly-stable part of parameter space, suggesting that secular processes  alone, while adiabatically changing the disc-to-halo and bulge-to-disc mass fractions and geometry, should not trigger bi-symmetric instabilities  leading to bar formation. 

These findings can be qualitatively explained either in an orbital alignment framework or a wave amplification one. In the former, the growing bulge (increasing $p$) component deflects the alignment of elongated orbits which could build up the bar,  while the lighter disc (decreasing $q$) weakens the relative torques such tumbling orbits collectively apply on each other, preventing bar growth \citep{Lynden-bell1979}. In the latter, the growth of the bulge generates an inner Lindblad resonance which absorbs swing amplified waves \citep{Goldreich1978}, breaking the amplification cycle towards bar growth. Conversely, the swing amplification is stronger in less dark matter dominated galaxies. The orbital formulation should also apply to the process of dissolving existing (non-linear) bar via cosmic gas infall, shepherded towards the centre by the bar, and building up a deflecting bulge.

The toy model  highlights an alternative summary statistic space, $(p, q)$ complementing that presented in Sec.~\ref{subsection:rotationcurves}, leading to an easily computed, physically motivated threshold for secular bar formation. It clearly does not allow to capture in full  the realm of complex processes relevant to bar formation in general.  For instance, beyond the aforementioned caveats, \cite{Aoki1979}'s formalism cannot rule out induced bar growth e.g. via strong tidal perturbations nor does it account for the impact of a live halo.    Yet, within that framework, we can conclude that the measured bulge size and disc mass fractions  explain the lack of bars in \newh. 

In conclusion,  while the \newh simulation has been shown to reproduce several key properties that define galaxies in reasonable agreement with observations,  the sub-grid physics encoded in \newh   \citep[and more generally in recent cosmological simulations  resolving discs scale heights down to low redshifts, e.g. ][]{Hopkins_2018, Agertz_2020} seems to either lack resolution or induce a galactic assembly history statistically incompatible with secular bar formation. This `bar problem' will need to be addressed in future work.

\section{False Positives and Galaxy Substructure}
\label{sec:technicalities}

In Section \ref{Bar_Fractions}, we suggested that our two bar detection methods may produce `false positives' whereby bar-like properties specific to each method are incorrectly classified as a bar and that therefore, a concurrence between the two methods is needed in order to confirm a galaxy as barred. In order to decipher the causes behind any biases in our bar detection methods, we will now inspect our galaxies visually as well as comparing detection methods on a galaxy-galaxy basis. We will then study further the impact that galaxy substructures have on bar detection in these methods.

\subsection{ A visual inspection on a galaxy-galaxy basis}
\label{visual inspection results}

\begin{figure*}
 \centering
 \includegraphics[width=\textwidth]{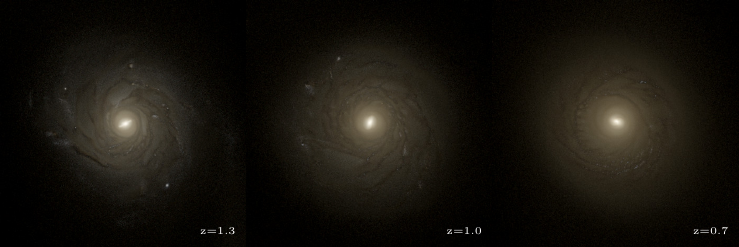}
 \caption{Mock observations of the only  detected and visually confirmed barred galaxy at redshift $z=1.3$ (\textit{left}), $z=1.0$ (\textit{middle}) and $z=0.7$ (\textit{right}). Mock images are produced  face-on (with respect to the stellar angular momentum of the galaxy) in SDSS g-r-i bands using the SKIRT9 code \protect\citep{CampsBaes_2020} that computes radiative transfer effects based on the properties and positions of the stars and the dusty gas, assuming a dust fraction of 0.4 following \protect\cite{Saftly_2015}. Each panel has a size of 23.5 kpc across and the stellar mass of the galaxy is $\sim 10^{11}$ \msun, with only a very weak evolution across this redshift range. 
}
 \label{fig:barred_gal_z}
\end{figure*}

An initial and rudimentary visual inspection of all the galaxies in our sample was carried out on the stellar mass surface density (face-on) projections of each galaxy. From this we find only \textit{one} clear, elongated bar structure across the entire sample. This galaxy is presented in Figures~\ref{fig:Barred_Galaxy} (stellar density) and~\ref{fig:barred_gal_z} (mock images). It is also apparent that the galaxies in our samples are highly perturbed, even at \(z = 0.25\), making any visual detection of smaller, more perturbed bars challenging. When applying the surface density method, we found that only $\sim$2\% of galaxies hosted an observable bar (or $\sim$ 1\% for "strong" bars) at redshifts \(z = 0.25\), 0.5 and 0.7 with this bar fraction increasing to 6 - 8\% at redshifts \(z = 1.0\) and 1.3 (see Table \ref{tab:Bar_Fractions}). This highlights the complications involved in attempting to detect bars visually at high redshift due to the largely perturbed nature of these galaxies. Example grids of the most massive \newh galaxies at \(z = 1.3\) and 0.25 are shown in Appendix \ref{fig:Galaxy_Grid_1_3} and \ref{fig:Galaxy_Grid_0_25} respectively to highlight their perturbed nature. When applying the velocity method to detect barred galaxies, we find there to be similar numbers of strong bars but slightly more observable bars in these samples. 

\begin{table}
\caption{The final results of our bar search analysis in terms of bar fractions calibrated between the two methods whereby a barred galaxy is confirmed as such if said galaxy found to be barred in by both methods. The bar fractions (and number of barred galaxies) are shown for redshifts \(z = 1.3, 1.0, 0.7, 0.5\) and \(0.25\). The confidence intervals presented have been estimated using the (Bayesian) beta distribution quantile technique \citep{Cameron_2011} as in Table \ref{tab:Bar_Fractions}.}
\centering
\begin{tabular}{ccc}
\cline{2-3}
         & \multicolumn{2}{c}{\begin{tabular}[c]{@{}c@{}}Final Bar Fractions\\ (Number of Bars)\end{tabular}}                                                                \\ \hline
Redshift & Strong Bars                                         & Observable Bars                                                                        \\ \hline
1.3      & \begin{tabular}[c]{@{}c@{}}0.020$_{-0.005}^{+0.012}$\\ (6)\end{tabular} & \begin{tabular}[c]{@{}c@{}}0.070$_{-0.012}^{+0.018}$\\ (21)\end{tabular}  \\
1.0      & \begin{tabular}[c]{@{}c@{}}0.019$_{-0.005}^{+0.012}$\\ (5)\end{tabular} & \begin{tabular}[c]{@{}c@{}}0.035$_{-0.008}^{+0.015}$\\ (9)\end{tabular}   \\
0.7      & \begin{tabular}[c]{@{}c@{}}0.004$_{-0.001}^{+0.010}$\\ (1)\end{tabular} & \begin{tabular}[c]{@{}c@{}}0.018$_{-0.005}^{+0.013}$\\ (4)\end{tabular}   \\
0.5      & \begin{tabular}[c]{@{}c@{}}0.009$_{-0.003}^{+0.012}$\\ (2)\end{tabular} & \begin{tabular}[c]{@{}c@{}}0.023$_{-0.006}^{+0.015}$\\ (5)\end{tabular}   \\
0.25     & \begin{tabular}[c]{@{}c@{}}0.005$_{-0.002}^{+0.012}$\\ (1)\end{tabular} & \begin{tabular}[c]{@{}c@{}}0.011$_{-0.003}^{+0.014}$\\ (2)\end{tabular}   
\end{tabular}
\label{tab:Final_Bar_Fractions}
\end{table}

Disparities in the specific galaxies selected as barred by each method suggest that in each of these methods, there is some bias producing 'false positives'. By comparing between the results in Tables \ref{tab:Bar_Fractions} and \ref{tab:Final_Bar_Fractions} (described below), we see that across the full sample of galaxies, 1-2\% of galaxies produce false positives in the surface density method and between 2-10\% of galaxies produce false positives in the velocity method.

By inspecting these supposed barred galaxies in detail, both visually and with plots specific to the two bar detection methods, we find that the surface density method is actually picking out regions of constant phase in galaxies - either due to the galaxies being rather perturbed or elongated due to galaxy-galaxy interactions or mergers or due to disordered structure in the central regions of the galaxy - where the stellar orbits are not actually aligned and trapped within a bar and so are not detected by the velocity method (which will be discussed more in Section \ref{impact_of_substructures} below). The velocity method also has its own bias whereby it finds some regions of aligned orbits as part of a `bar region', but the very same region does not have a Fourier phase, $\Phi_{2}$, constant enough to be considered a bar by the surface density method. 

This disagreement highlights strongly the difficulties in detecting bars, and the importance of cross-checking methods. Both detection methods give distinct false positives, as has become apparent in this study, due to the absence of  many strong, elongated bars. For a robust detection, one can only confirm a galaxy as barred if both methods detect it as such. This cross-checking of two different detection methods (potentially coupled with a visual inspection - although less applicable at the low-mass end) is necessary then for studies of barred galaxies in the future. 

By saying that, to confirm a galaxy as barred, the galaxy needs to be detected by both methods, we calculate final bar fractions as a result of concurrences between the methods and these are presented in Table~\ref{tab:Final_Bar_Fractions}. We present in Appendix~\ref{sec:all_barred_galaxies}, for the reader's own visual inspection, face-on surface density projections of all the galaxies we confirm as barred. We stress here that our confirmation of a bar is based solely on agreements between the methods and that in this and the previous section, our `barred' galaxies refers to these theoretical or putative bars. These `bars' are rather perturbed and any visual inspection or observation would not be quick to suggest that these are, by any means, clear or certain bars. We can also see that, towards the low-mass regime, i.e. \(\log (\mstar/\msun) < 8.0\), it becomes increasingly difficult to identify any structure in the galaxy visually. In this regime, bar detection can only rely on our bar detection methods as any visual inspection will struggle to probe this mass range. These bars are often short-lived (only detected at their given redshift) and our overall finding of no clear, elongated bars found in \newh remains consistent. The only exception is the galaxy presented in Figures~\ref{fig:Barred_Galaxy} (stellar density) and~\ref{fig:barred_gal_z} (mock images), which remains strongly barred from \(z = 1.3\) - 0.7 (see the first galaxy at each \(z = 1.3\), 1.0 and 0.7 in Figure~\ref{fig:barred_galaxy_grid}). This is a fairly short (bar length $\sim$ 1.3 kpc), central bar but is the clearest and strongest detection of a barred galaxy we find and is the only bar that remains over multiple redshifts. The galaxy is also the most massive disc galaxy in the sample at \(z=1.3\), \(\log (\mstar/\msun)=11.04\).

\subsection{The impact of substructures on detection}
\label{impact_of_substructures}

Let us finally examine in more  detail one bias that was brought up from the surface density method. That is disordered or clumpy substructure across the central regions of galaxies being detected as a bar. The galaxy identification algorithm AdaptaHOP as used in \newh removes the majority of these substructures. However, the HOP identification algorithm keeps all these substructures and this is essentially how a galaxy would be analysed in an observational study if star forming clumps are not removed from these visual inspections. Because of this, and the fact that we know our detection method can produce false positives as a direct consequence of these clumps, let us study the effect of these on our measurements by studying galaxies identified using the HOP algorithm.

In order to determine the effects of these substructures in our bar detection method, we ran our analysis again but this time on galaxies identified using the HOP algorithm and compared to the results using AdaptaHOP. We found that there was very little difference between the HOP and the AdaptaHOP galaxies when using the velocity method at any redshift. Focusing on the surface density method, at the lower redshifts, there was very little difference in the bar fractions and in those galaxies selected as barred between the two galaxy identification algorithms. However, at \(z = 0.7\) and higher, where galaxies tend to be more clumpy and perturbed \citep[see][and their Figures 4 and 5 for a quantitative measure of morphological disturbances at different redshifts]{Martin2021}, we found an increase in the bar fractions with more galaxies being detected as barred. This allows us to quantify the effect that including galaxy substructures has on measurements of the bar fraction via the surface density method. These HOP bar fractions are shown in Table \ref{tab:HOP_Bar_Fractions} for \(z=0.7\) along with results of some previous studies for comparison. 

\begin{figure*}
    \centering
    \includegraphics[width=\textwidth]{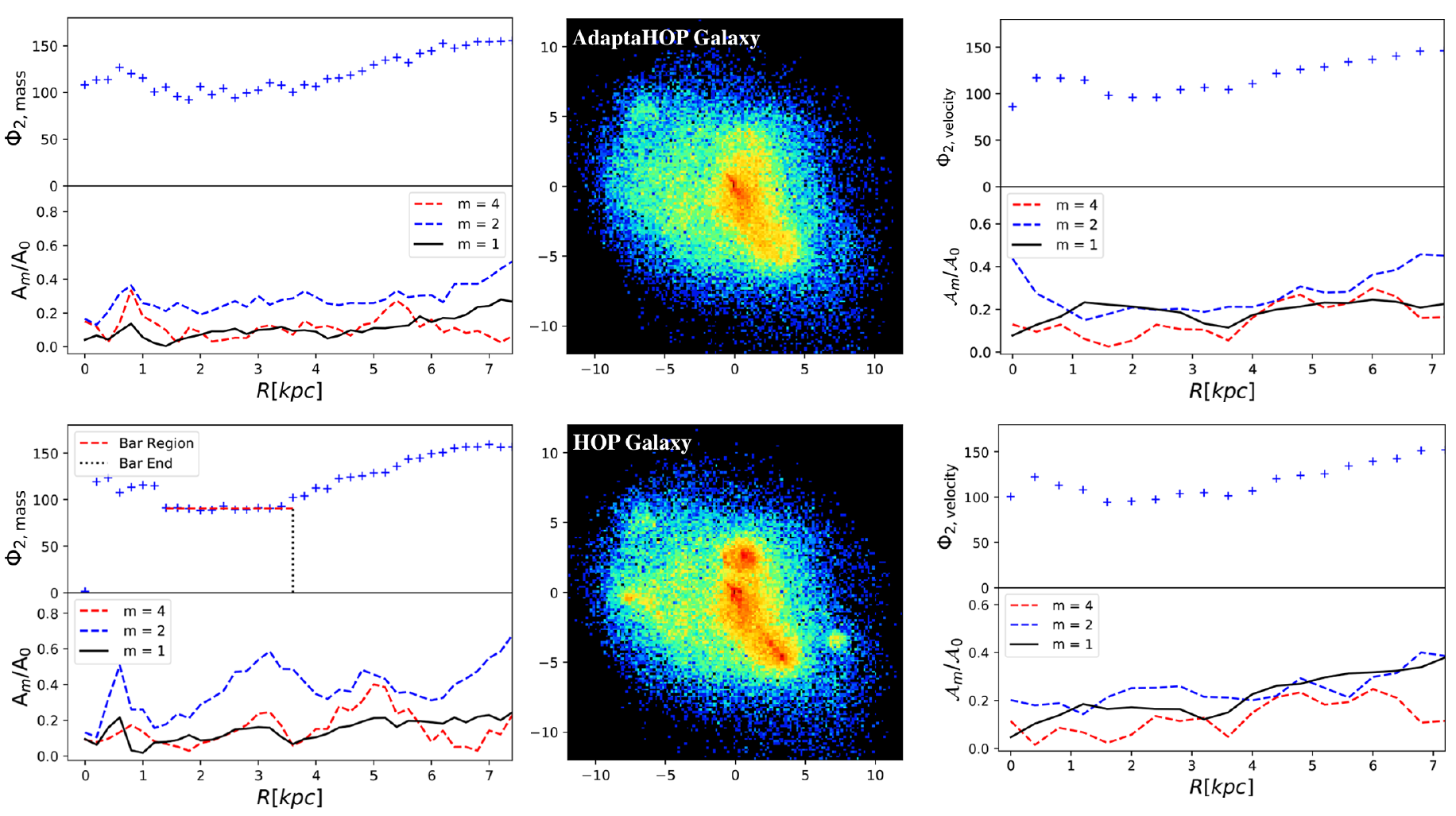}
    \caption{An example of a \newh galaxy at \(z = 0.7\) that initially wasn't found to be barred in either method when the galaxy was selected using the AdaptaHOP algorithm. However, when selecting the galaxy using the HOP algorithm so as to include substructure, this galaxy was found to be strongly barred in the surface density method. Stellar density maps are shown on the middle panels along with the corresponding Fourier phase and amplitudes versus radial distance plots for the surface density method (left) and for the velocity method (right). The top panels show the galaxy as selected using the AdaptaHOP identification algorithm with the galaxy selected using the HOP algorithm so as to include the substructure shown in the bottom panels. The surface density method's bar search procedure has detected a region of roughly constant phase across the centre of the face on projection of the HOP galaxy. As this has only been detected when substructures have been included, we can conclude that this is indeed not a barred galaxy, but is in fact a result of aligned clumps across the central regions of the galaxy. This highlights the caution required when observing bars visually at high redshift without the ability to perform a secondary, velocity-based bar detection method.}
    \label{fig:AdaptaHOP_Vs_HOP}
\end{figure*}

As this surface density method relies on detected regions of roughly constant Fourier phase, $\Phi_{2}$, we see that bars can be identified in a galaxy as a direct consequence of aligned clumps across the central region of a galaxy. As an example, we show in Figure \ref{fig:AdaptaHOP_Vs_HOP} a clumpy galaxy at $z=0.7$ in which the surface density method detects a bar only when analysing the galaxy extracted using the HOP algorithm\footnote{We also examined the dark matter map of this galaxy and we confirm that these `clumps' have no dark matter counterpart and therefore are real stellar clumps and not small, overlapping satellites.}. This highlights the potential problem with observing bars visually at high redshift, without the ability to perform a secondary, bar detection method. As the substructure of a galaxy cannot be easily removed from the observed images of galaxies, and given that galaxies at high \(z\) may be very perturbed, it is important to take into account a possible contamination by aligned stellar clumps that could be misidentified as bars.

\begin{table*}
\caption{Table presenting the strong and observable bar fractions at redshift \(z = 0.7\) as obtained through the surface density method using the HOP galaxy selection algorithm alongside the results of one numerical and two observational studies. The definition of a strong bar is broadly in agreement across all of these works. One can see that the HOP results obtained here have increased by 1-3\% compared to the previous AdaptaHOP results. This indicates that the presence of substructures may impact the bar identification and highlights the importance of carefully accounting for this effect, in particular at high redshift. The confidence intervals presented have been estimated using the (Bayesian) beta distribution quantile technique \protect\citep{Cameron_2011}.}
\centering
\begin{tabular}{ccccc}
\hline
Type of Study & Strong Bar Fraction & Observable Bar Fraction & Reference \\
\hline
Simulation    & $0.018_{-0.005}^{+0.014}$   &  $0.049_{-0.010}^{+0.019}$  & This Work \\[1pt]
Simulation    & $\sim$ 0.1          & $\sim$ 0.2  &  \citet{Kraljic_2012}   \\
Observational & $\sim$ 0.1          & -                  & \citet{Melvin_2014}    \\
Observational & $\sim$ 0.1          & $\sim$ 0.25       &  \citet{Sheth_2008}\\
\hline
\end{tabular}
\label{tab:HOP_Bar_Fractions}
\end{table*}

To summarise, we observe an increase in bar fraction measured via the surface density method when analysing HOP galaxies. This is due to the detection of clumpy structure aligned within the central regions of a galaxy. Thus we caution that, without the ability to remove substructures or to easily invoke the secondary velocity method on observations (without integral field spectroscopy), our finding confirms that clumps may introduce some bias in the measurement of the bar fraction at high redshift, and highlights the importance of accounting for possible contamination by clumps.

\section{The low bar fraction in context}
\label{Discussion}

In this section, we compare in more detail our measurements of the bar fraction to observations (Section~\ref{subsec:observationalcomparison}) and simulations (Section~\ref{subsec:simulationcomparison}), before discussing lessons learned for measuring bar fractions in both observations and simulations 
in Section~\ref{subsec:lessons}.

\subsection{Bar fractions measured from observations}
\label{subsec:observationalcomparison}

While there is a general consensus that the fraction of barred galaxies in the local Universe is high \citep[30\%-70\% depending on the bar classification method, the bar strength, the observed wave-bands and galaxy selection criteria][]{Eskridge_2000,Whyte_2002,Laurikainen_2004,Menendez-Delmestre2007,Marinova_Jogee_2007,Barazza_2008,Sheth_2008,Aguerri_2009,Nair_2010a,masters_2011,Masters_2012,Melvin_2014,Diaz_2016}, its stellar mass dependence is much more debated.

\citet{Erwin_2018} found, in the nearby ($z<0.01$) sample of spiral galaxies in the S$^4$G survey, that the barred fraction ($f_{\rm bar}$) increases steeply from $\sim$ 20\% at very low stellar mass ($\sim$ 10$^8$ \msun), reaching a maximum of $\sim$ 76\% at \mstar $\sim$ 10$^{9.7}$ \msun before declining to 50\% at \mstar $\sim $10$^{10-10.5}$ \msun and $\sim$ 40\% at \mstar $\sim$ 10$^{11}$ \msun. These results are in broad agreement with another study of S$^4$G galaxies \citep[][]{Diaz_2016} reporting an increasing bar fraction from about 0\%-15\% (depending on the adopted method) for low mass (10$^{8-8.5}$ \msun) galaxies to 40\%-60\% at $\sim$ 10$^{9.5}$ \msun. \cite{Diaz_2016} found that the fraction of bars then stays roughly constant out to 10$^{11}$ \msun. As suggested by \citet{Erwin_2018}, this difference at high stellar masses can be attributed to selection criteria, mainly by the presence of S0 galaxies in the sample analysed by \cite{Diaz_2016} producing a flat $f_{\rm bar}$-\mstar trend. These measurements are in disagreement with most of the SDSS-based studies \cite[e.g.][]{Masters_2012,Oh_2012,Melvin_2014} which report $f_{\rm bar}$ that increases strongly with stellar masses (for \msun $> 10^{10}$ \msun typically). As shown by \cite{Erwin_2018}, this is likely due to the difficulty of SDSS-based studies to identify smaller bars at low stellar mass ($< 10^{10}\msun$). The resulting underestimate of the bar fraction at these stellar masses leads to their incapability in tracing a peak in bar fraction at $\sim10^{9.7}\msun$. A number of near infrared studies \cite[e.g.][]{Diaz_2016,Erwin_2018} also find that bar sizes depend on galaxy stellar mass -- in the high mass regime, $\mstar > 10^{10.2} \msun$, bar length is found to increase with galaxy mass while lower mass galaxies tend to have bar lengths roughly constant at $\sim$ 1.5 kpc. This could suggest a limit on the size of bars one can resolve visually in these observational studies of lower mass galaxies with one possible explanation being that short bars could be missed when they coexist with a massive bulge \citep{Zhao_2020_Barred_IllustrisTNG}.

As opposed to the nearby Universe, measuring the bar fraction at higher redshifts proves to be increasingly difficult due to a lack of spatial resolution and band-shifting \cite[see][]{Sheth_2003}. But thanks to the high resolution deep optical and near-infrared data, it is today established that the fraction of barred galaxies decreases with increasing redshift \citep[e.g.][]{Sheth_2008,Cameron_2010,Melvin_2014,Simmons_2014}. These high redshift studies focus primarily on stellar masses above 10$^{10}$ \msun and find that $f_{\rm bar}$ i) increases with stellar mass from 15\% (5\%) for masses  $\sim$10$^{10.3}$ \msun at z$\sim$ 0.5 (0.9) to $\sim$ 35\% (10\%) at 10$^{10.9}$ \msun at z$\sim$ 0.5 (0.9); ii) increases with time (from $\sim$ 40\% at z$\sim$0.25 to $\sim$ 10\% at z$\sim$1); or stays roughly constant for strong bars. Based on these observations, we expect there to be some evolution in $f_{\rm bar}$ with both mass and redshift in the high mass regime, and that the bar fraction in the low mass regime is low.

Clearly, the low bar fractions measured in \newh (see Section~\ref{Bar_Fractions}) do not match the local sample of galaxies. The comparison is less obvious at higher redshift, but as our results are consistent with few bars in \newh at any redshift, our findings are in tension with higher redshift observations as well. We do stress that we are probing, on average, a lower mass regime than Milky Way-mass local galaxies, or the mass regimes that are currently accessible in higher redshift observations (typically \mstar~>~10$^{10}$~\msun). However, applying a similar stellar mass cut, i.e. \mstar > 10$^{10}$ \msun, so as to make a more direct comparison to observations, does not change our conclusions. The fraction of strong and observable bars in \newh galaxies above this mass cut is found to decrease from $f_{\rm bar}$~=~0.167 at $z\sim$ 1.3 down to zero at $z\sim$ 0.25 (see Tables \ref{tab:Bar_Fractions_Mass_Cut} and \ref{tab:Final_Bar_Fractions_Mass_Cut}), in disagreement with both observed $f_{\rm bar}$ at low redshifts and its trend with cosmic time. Future space and ground-based facilities will be crucial for constraining bar fractions at increasing redshift.

\subsection{Bar fractions measured in simulations}
\label{subsec:simulationcomparison}

Large-scale cosmological simulations provide a powerful tool in studying the formation and evolution of bars. Recently, \citet{Rosas_Guevara_2020}, \citet{Zhou_2020} \& \citet{Zhao_2020_Barred_IllustrisTNG} present studies of barred galaxies in the \tnghundred simulation. \citet{Rosas_Guevara_2020} find a bar fraction of $\sim$40\% at \(z = 0\) among the disc galaxies with stellar masses in the range 10$^{10.4-11}$ \msun. The fraction of barred galaxies is found to increase with stellar mass, more clearly for strong than for weak bars. The total bar fraction increases from 25\% for the lowest to 75\% for the highest stellar mass bin in their sample. They also find that strong bars tend to form in galaxies where a prominent disc component was established early (between \(0.5 < z < 1.5\)) with more massive disc galaxies tending to host older bars whereas unbarred galaxies tended to form their disc components at later times. \citet{Zhou_2020} present a comparison of the properties of bars between the \illustris and the \tnghundred simulations focusing mainly on the discrepancies between the two simulations as a result of star formation and AGN feedback. They find that, at \(z=0\), the fraction of bars among disc galaxies with  stellar mass \mstar $> 10^{10.5}$ \msun in \tnghundred is 55\% whereas in \illustris it is 8.9\%. The bar fraction increases with stellar mass in both simulations. In \tnghundred, it increases rapidly from 0\% in the lowest stellar mass bin (10$^{10.25-10.33}$ \msun), to 30\% in the intermediate mass bin (10$^{10.50-10.58}$ \msun), and to 50\% in the range 10$^{10.66-11.25}$ \msun. In \illustris, the bar fraction increases gradually from 0\% in the stellar mass bin 10$^{10.50-10.58}$ \msun, to 10\% for galaxies with \mstar $\sim 10^{11}$ \msun, and then grows rapidly to 30\%--40\% for galaxies with \mstar  $> 10^{11.25}$ \msun, in agreement with \cite{Peschken2019}, analysing the same simulation. The differences between these two simulations is suggested to be due to the combination of more effective stellar and AGN feedback in \tnghundred which cause massive galaxies to have lower gas fractions at low redshifts, therefore aiding bar formation. Moreover, the work of by \cite{Athanassoula_2013} indicates that gas-rich galaxies may indeed experience a delay in bar formation. By comparing galaxies directly between the \illustris and the \tnghundred simulations, \cite{Zhou_2020} find that a galaxy's morphology (and whether or not it is barred) at \(z=0\) is dependent not only on the internal baryonic physics, but also on the environment - whether a galaxy experiences a merger, a flyby or solely secular evolution - which is often unpredictable. \citet{Zhao_2020_Barred_IllustrisTNG} also study bars in the \tnghundred simulation, selecting galaxies with stellar masses \mstar $\geq$ 10$^{10}$ \msun. At $z=0$ the bar fraction is found to increase with stellar mass from $\sim$ 5\% at stellar mass of 10$^{10}$ \msun up to $\sim$ 75\% at \mstar $\geq 10^{10.6}$ \msun, in disagreement with the observed almost constant bar fraction of 50\%--60\% over the same stellar mass range in the nearby Universe \citep[][]{Erwin_2018}. When tracing the progenitors of $z=0$ massive galaxies (with \mstar $> 10^{10.6}$ \msun), they find an increasing bar fraction from 25\% at $z=1$ to 63\% at $z=0$. Applying a constant mass cut of \mstar $> 10^{10.6}$ \msun in the redshift range $z=0-1$ reveals instead a nearly constant bar fraction of 60\%, while observations using a similar sample selection show a dramatic reduction of bars across the same cosmic epoch. They suggest that this discrepancy is due to the resolution of \tnghundred and its inability to resolve bars with radii $\leq 1.4$ kpc, as well as alluding to the possibility that observations may fail to identify many of these short bars at high redshifts. 

\cite{Rosas-Guevara_2021} study the evolution of the barred galaxy population in the recent \tngfifty simulation. They focus on galaxies with masses \mstar > 10$^{10}$ \msun and find a bar fraction of $\sim30\%$ at \(z=0\) which evolves mildly to being above $\sim 40\%$ at \(0.5 < z < 3\). \tngfifty is comparable to \newh in terms of resolutions and volume and reproduces fairly well the cosmological evolution of the main properties of disc galaxies \citep{Pillepich_2019,Nelson_2019}. While this high-resolution hydrodynamical simulation produces a bar fraction in rough agreement with observations at \(z\) = 0, they fail to reproduce the observed declining bar fraction at increasing \(z\) (> 0.5).

The redshift evolution of barred galaxies have also been addressed by \cite{Peschken2019} in the \illustris simulation, by analysing the evolution of high mass (\mstar > 10$^{10.9}$ \msun) disc galaxies. 21\% of these are found to be barred at $z=0$, while this fraction increases slightly with redshift. Most of these bars are triggered by external perturbations such as mergers or flybys, many of which disappear during secular evolution, leading to a lower bar fraction at present time compared to observations. Finally \cite{Eagle_Bars_2017} examined disc galaxies with stellar masses in the range 10$^{10.6-11}$ \msun in the \eagle simulations at $z=0$, finding a total fraction bars of 40\%, with 20\% being strongly barred. 

In summary, at $z=0$ all large-scale cosmological simulations seem to overproduce bars at high masses (\mstar $\gtrsim$ 10$^{10.5}$ \msun) and tend to suppress their formation at low masses (in the stellar mass range \mstar = $10^{10-10.5}$ \msun) compared to observations of the nearby Universe \citep[][]{Erwin_2018}. These simulations also fail to reproduce the declining bar fraction with redshift, seen in most observations.  

We stress again that the majority of previous studies, both observational (in particular at high redshift) and simulated, study larger mass galaxies due to the limited spatial resolution. Typically, they focus on galaxies with stellar masses \mstar $> 10^{10-10.5}$ \msun, whereas the number statistics above this galaxy stellar mass in the \newh simulation is low \citep[see Table 1 of][for more details on galaxy masses within \newh]{Dubois_2020}.  Therefore, any comparison made to the literature in this section is not direct as this work presents an initial study into bars in low mass galaxies down to redshift \(z = 0.25\) owing to the higher spatial resolution of \newh. However, as discussed in Section~\ref{subsec:observationalcomparison}, when we limit our sample to galaxies with stellar masses \mstar $> 10^{10}$ \msun, disagreement with these studies becomes even more striking, in particular at low redshift ($z=0.25$), where the bar fraction decreases to zero (see Tables \ref{tab:Bar_Fractions_Mass_Cut} and \ref{tab:Final_Bar_Fractions_Mass_Cut}). Stabilisation of these massive disks in \newh against the formation of bars was further discussed in Section~\ref{sec:lowbarfractions}.

Future work is needed to measure bars homogeneously  between simulations, for a homogeneous sample of simulated galaxies. We advocate  using {\sl both} methods employed in this work, which will eliminate spurious detections.

\subsection{What are the lessons for bar fraction measurements?}
\label{subsec:lessons}

The initial visual inspections of all galaxies extracted from the simulation showed that there was only one clear, strong, extended bar and that this bar was observed from \(z = 1.3\) to \(z = 0.7\) only. When applying a galaxy-by-galaxy comparison, we found disparities between the two methods: a detailed inspection revealed biases specific to each method (see Section \ref{visual inspection results}) resulting in both methods giving false positives. This cross-checking of detection methods led to a final bar fractions of $f_{\rm bar}$ = 0.070 at  $z\sim$ 1.3 which decreases with decreasing redshift down to $f_{\rm bar}$ = 0.011 at  $z\sim$ 0.25 (see Table \ref{tab:Final_Bar_Fractions} for our full results), in disagreement with the majority of previous studies. Limiting our sample to galaxies with \mstar > 10$^{10}$ \msun resulted in final bar fractions decreasing from $f_{\rm bar}$ = 0.167 at $z\sim$ 1.3 to zero at $z\sim$ 0.25 (see Tables \ref{tab:Bar_Fractions_Mass_Cut} and \ref{tab:Final_Bar_Fractions_Mass_Cut}), in even stronger disagreement with previous studies at low redshift.

While these results may suggest a failure for the simulation to reproduce a striking feature of the observed Universe, it does highlight possible biases of observational studies of barred galaxies at lower mass and/or high \(z\) galaxies, and highlight the importance of robust detection methods. 

Our initial bar search criteria required a potential bar region to be at least 1 kpc in length before classifying the galaxy as barred. By reducing this criterion to 0.5 kpc, we found virtually no change in our results. This, along with the perturbed nature and lack of visual clarity in the bars we detect, as well as bars being detected in lower mass galaxies all point to the increasing difficulty with which to confirm a bar visually as bars become shorter and more perturbed. These shorter bars are increasingly difficult to confirm as barred, owing to several reasons:
\begin{enumerate}
    \item When approaching the lower mass regime and/or when trying to resolve much shorter bars, we are approaching the limit of application of the Fourier analysis to a surface density or a velocity field.
    \item In the lower mass and short-bar regimes, the susceptibility of the two detection methods to noise in the determination of \(m = 2\) Fourier amplitude becomes a problem with the scale of the noisy regions becoming comparable to possible bar regions.
    \item Both methods we employ are susceptible to producing false positives (e.g. regions of constant Fourier phase $\Phi_{2}$). False positives are increasingly common if the minimum length criterion is reduced.
    \item  It is difficult to confirm  short or weak bars visually, especially if these short bars coexist with a massive bulge or in lower mass galaxies where the structure of the galaxy.
\end{enumerate}

Given our identified  biases, which may be present in both observations and simulations, we highlight the difficulties involved in current observational estimates of bar fraction - whereby caution must be taken to exclude aligned stellar clumps or that shorter bars are unresolved - at high redshift and/or in the low-mass regime.

\section{Summary}
\label{Summary}

We studied the redshift evolution of the bar fractions in galaxies extracted from the \newh simulation at redshifts \(z = 1.3\), 1.0, 0.7, 0.5 and 0.25. We selected 299, 260, 224, 221 and 183 disc galaxies at each respective redshift using the stellar kinematics of a galaxy as a proxy to infer its morphology. Galaxy masses ranged $10^{7.25}\,\msun \leq M_{\star} \leq 10^{11.4}\,\msun$, with only a handful (13 at \(z = 1.3\) and 21 at \(z = 0.25\)) higher mass ($M_{\star} \geq 10^{10} \,\msun$) galaxies in our sample.
We implemented two bar detection methods - one based on the harmonic decomposition into Fourier components of the stellar surface density profile of the galaxy and the second based on the harmonic decomposition into Fourier components of the tangential velocity field of the galaxy - in order to provide a more robust measurement of the bar fraction. We then analysed galaxy rotation curves and growth rates to gain an insight into the measured low bar fractions and then studied biases that arose in the two methods and the impact of substructure in galaxies on  bar fractions.
The main results of our paper are as follows:
\begin{enumerate}
   \item A cursory dynamical analysis indicates that galaxies in \newh may not reach thresholds  in stellar content and distribution for bar formation. At lower masses
   (\mstar<10$^{10}$\msun),
   galaxies appear to be too dominated by dark matter relative to stellar content. At higher masses
   (\mstar>10$^{10}$\msun),
   galaxies appear to be stabilised by the presence of a central bulge.
    \item We confirm barred galaxies as such by requiring detections in both methods. This results in final observable bar fractions of $f_{\rm bar}$ = 0.070$_{\scalebox{.5}{-0.012}}^{+\scalebox{.5}{0.018}}$ at z $\sim$ 1.3 which decreases down to $f_{\rm bar}$ = 0.011$_{\scalebox{.5}{-0.003}}^{+\scalebox{.5}{0.014}}$ at z $\sim$ 0.25. This is in disagreement with the majority of past studies, both observational and numerical, namely large-scale cosmological simulations. Visual inspections of candidate barred galaxies revealed biases specific to each method, both  giving false positives. 
    \item The measured bar fractions of disc-dominated galaxies  in \newh  are too low compared to observational results reported in the literature. The most massive galaxies tend to host the weakest bars; there is little to no redshift evolution of the bar fraction, which all point to a challenge for current state-of-the-art galaxy formation models.  
\end{enumerate}

This work shows that the assembly history of the \newh galaxies appears to quench the formation of any strong bars. It also highlights possible biases for observational studies of barred galaxies at lower mass and/or high \(z\) galaxies, and highlights the importance of employing robust detection methods.

However, we stress that we are probing, on average, a lower mass regime than Milky Way mass local galaxies, or the mass regimes that are currently accessible to higher redshift observations. 
Future space-based observatories, such as with the James Webb Space Telescope, the Nancy Grace Roman Telescope, and other planned ground-based facilities will be crucial to constraining bar fractions at increasing redshift.

Beyond the scope of this paper, it would be interesting to track individual galaxies through cosmic time and  study in details the  evolution of their dark matter, gas and stellar components. This would help constraining further the parameters space favouring the formation of bars and eventually pin down {\sl why} galaxies in \newh do not appear to form bars. This will be addressed in the future work.

\section*{Data availability}
The data underlying this article will be shared on reasonable request to the corresponding author.
The code of the linear stability analysis is available online at \href{https://github.com/KerwannTEP/LiRA}{https://github.com/KerwannTEP/LiRA}.

\section*{Acknowledgements}

We thank Oscar Agertz and Florent Renaud for stimulating discussions.
This work was granted access to the HPC resources of CINES under the allocations  c2016047637, A0020407637 and A0070402192 by Genci, KSC-2017-G2-0003 by KISTI, and as a “Grand Challenge” project granted by GENCI on the AMD Rome extension of the Joliot Curie supercomputer at TGCC.  This research is part of ANR Segal ANR-19-CE31-0017 (\href{http://secular-evolution.org}{http://secular-evolution.org}) and Horizon-UK projects. This work has made use of the Horizon cluster on which the simulation was post-processed, hosted by the Institut d'Astrophysique de Paris. We warmly thank S.~Rouberol for  running it smoothly. CP and KT thank J.B.~Fouvry for numerous feedback.  KK acknowledges support from the DEEPDIP project (ANR-19-CE31-0023). 
MSP acknowledges funding from a UK Science and Technology Facilities Council (STFC) Consolidated Grant.
S.K.Y. acknowledges support from the Korean National Research Foundation (NRF-2020R1A2C3003769). 
RJ acknowledges support from the Yonsei University Research Fund (Yonsei Frontier Lab. Young Researcher Supporting Program) of 2021 and by the Korean National Research Foundation (NRF-2020R1A2C3003769).
The supercomputing time for numerical simulation was kindly provided by KISTI (KSC-2017-G2-003), 
and large data transfer was supported by KREONET, which is managed and operated by KISTI. 




\bibliographystyle{mnras}
\bibliography{bib} 




\appendix

\section{Barred Galaxies}
\label{sec:all_barred_galaxies}

\begin{figure*}
    \centering
    \includegraphics[width=\textwidth]{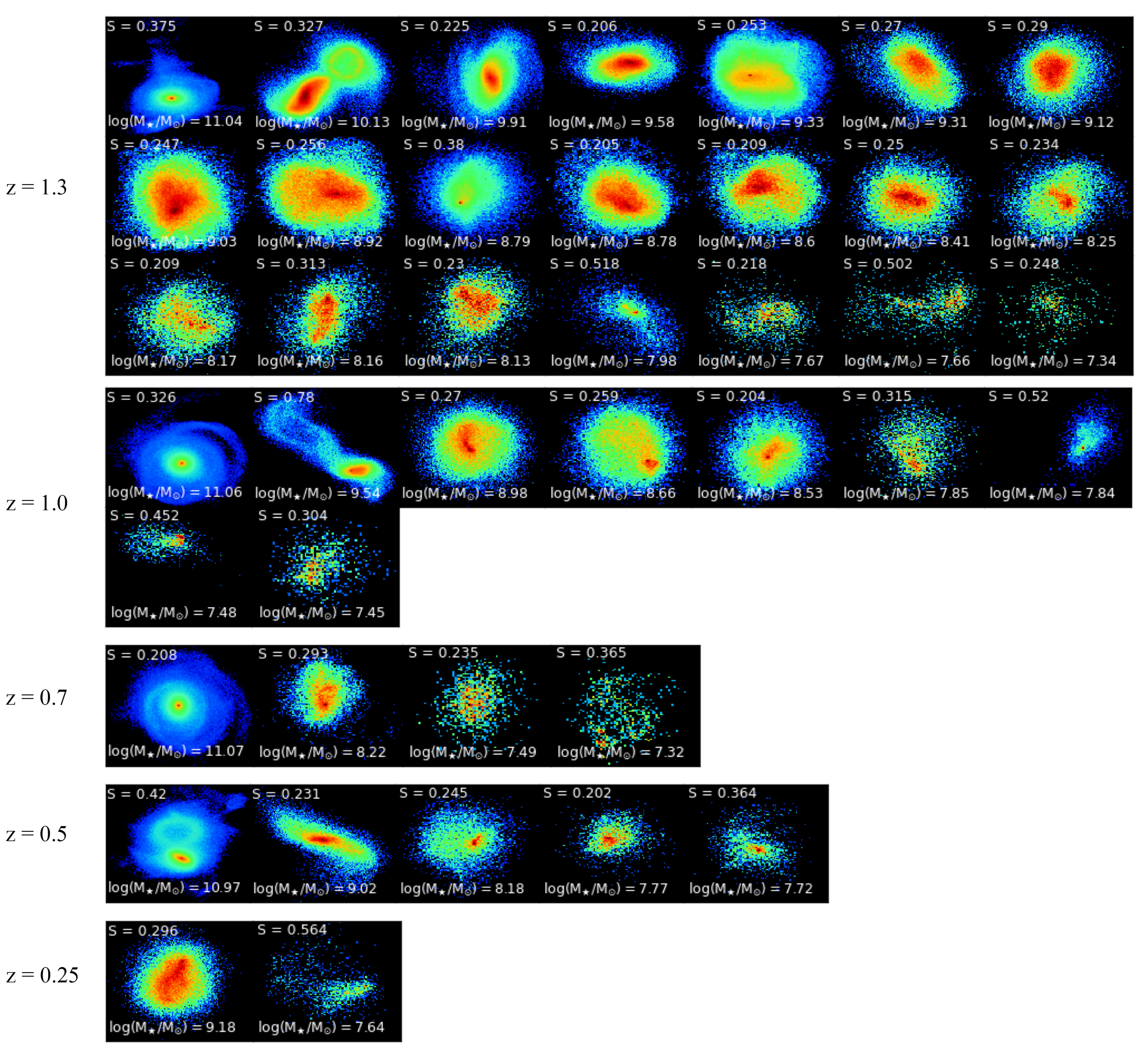}
    \caption{In this figure, we present, for the readers interest and discussion, the stellar surface density face-on projections of all the galaxies that we confirm as barred at each redshift. The strength of their bar is indicated in the top left corner, while their stellar mass is given at the bottom left. These barred galaxies are confirmed as such by requiring a bar detection in both detection methods applied. It is clear that there are no clear, large, elongated bars in this simulation and that the bars detected here are mostly small and/or rather perturbed. Confirmation of a bar is based solely on agreements between the methods and that our 'barred' galaxies refers to these theoretical or putative bars. All bars are short-lived (only detected at their given redshift) with one exception - that of the galaxy presented in Figure \ref{fig:Barred_Galaxy}, which remains strongly barred from \(z = 1.3\) - 0.7 and can is presented as the first galaxy at each \(z = 1.3\), 1.0 and 0.7 in this Figure.}
    \label{fig:barred_galaxy_grid}
\end{figure*}

As discussed in section \ref{visual inspection results}, to confirm a barred galaxy as such, we required that a bar be detected via both methods applied. This was in order to remove any false positives produced by either method. In Figure~\ref{fig:barred_galaxy_grid}, we present, for the readers interest and discussion, the surface density face-on projections of all the galaxies that we confirm as barred via this process at each redshift. It is clear that there are no clear, large, elongated bars in this simulation and that the bars detected here are mostly small and/or perturbed. 

We remind the reader that our confirmation of a bar requires agreement between the methods and that our `barred' galaxies refers to these theoretical or putative bars. These `bars' are perturbed and that any visual inspection would not be quick to suggest that these are, by any means, clear or certain bars. These are often short-lived (only detected at their given redshift) and that our overall finding - no clear, elongated bars found in \newh - remains consistent. The only exception is the galaxy presented in Figure \ref{fig:Barred_Galaxy}, which remains strongly barred from \(z = 1.3\) - 0.7 (see the first galaxy at each \(z = 1.3\), 1.0 and 0.7 in Figure \ref{fig:barred_galaxy_grid}). This is a fairly short (bar length $\sim$ 1.3kpc), central bar but is the clearest and strongest detection of a barred galaxy we find and is the only bar that remains over multiple redshifts. 

Some of the barred galaxies detected here are in the lower mass regime - towards the lower limit of the Fourier based methods where noise in the radial bins starts to become significant. These galaxies do not show clearly on these projections and suggest that in the very low mass range (below $\sim 5 \times 10^{7} \,\msun$), we are approaching the limit of bar detections via these methods in this simulation. These again further demonstrate the difficulties in observing bars at higher redshift and in the lower mass regime. 

\section{Perturbed Galaxies}

\begin{figure*}
    \centering
    \includegraphics[width=\textwidth]{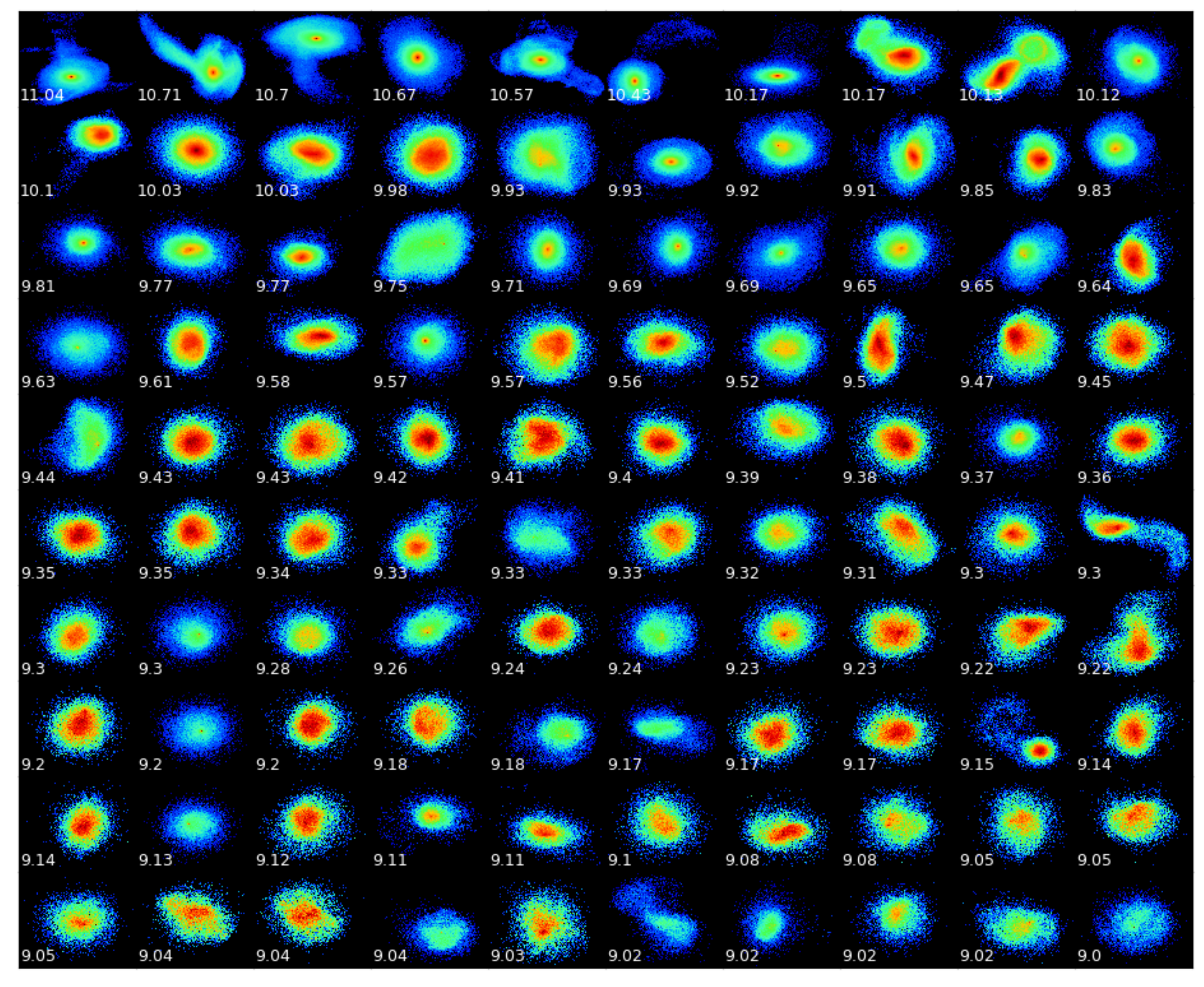}
    \caption{Initial visual inspection of the 100 most massive galaxies extracted from \newh at \(z = 1.3\) (the logarithm of the stellar mass of each galaxy is represented by the number on the given panel i.e. the top left galaxy has a stellar mass $\mstar = 10^{11.04} \msun$).
    It is clear that these galaxies are highly turbulent and that there are no clear, strong, elongated bars in any of the galaxies that could be seen visually at any of the redshifts. The surface density maps are plotted for $50 \times 50 \, \text{kpc}^{2}$ and the colour coding scale is logarithmic as in Figures \ref{fig:Barred_Galaxy} and \ref{fig:Unbarred_Galaxy}.
    }
    \label{fig:Galaxy_Grid_1_3}
\end{figure*}

\begin{figure*}
    \centering
    \includegraphics[width=\textwidth]{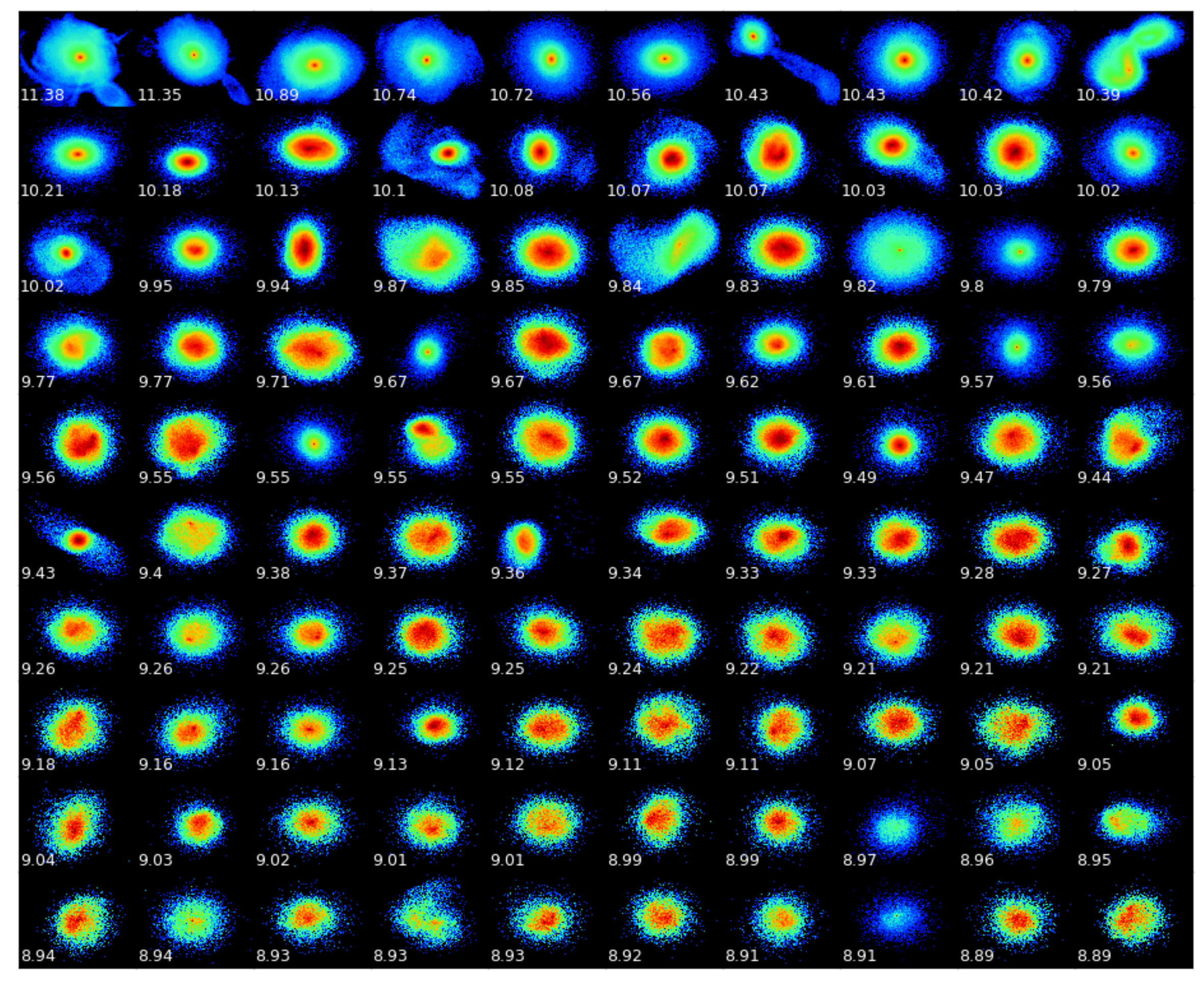}
    \caption{Initial visual inspection of the 100 most massive galaxies extracted from \newh at \(z = 0.25\) (the logarithm of the stellar mass of each galaxy is represented by the number on the given panel i.e. the top left galaxy has a stellar mass $\mstar = 10^{11.38} \msun$).
    One can still see that these galaxies are still highly turbulent highlighting the difficulty in identifying barred structure visually. The surface density maps are plotted for $50 \times 50 \, \text{kpc}^{2}$ and the colour coding scale is logarithmic as in Figures \ref{fig:Barred_Galaxy} and \ref{fig:Unbarred_Galaxy}.
    }
    \label{fig:Galaxy_Grid_0_25}
\end{figure*}

Figures \ref{fig:Galaxy_Grid_1_3} and \ref{fig:Galaxy_Grid_0_25} shows examples of initial visual inspections of the 100 most massive galaxies extracted from the \newh Simulation at \(z = 1.3\) and at \(z = 0.25\) (with masses ranging from $\sim 10^{9} \,\msun $ to $ \sim 10^{11} \,\msun$). These grids were used to perform visual inspections on all the galaxies in the samples at each redshift in order to determine the number of galaxies that may potentially be barred. However, almost none of these were clear, strong, elongated bars. It is also apparent that these galaxies are highly perturbed, even at $z = $ 0.25, making any visual detection of smaller, more perturbed bars challenging. These grids are presented here to highlight the complications involved in attempting to detect bars visually (especially at high redshift) due to the largely perturbed nature of these galaxies.

\section{Bar Properties in the Velocity Based Method}
\label{Appendix: velocity based method}

\begin{figure}
    \centering
    \includegraphics[width=0.5\textwidth]{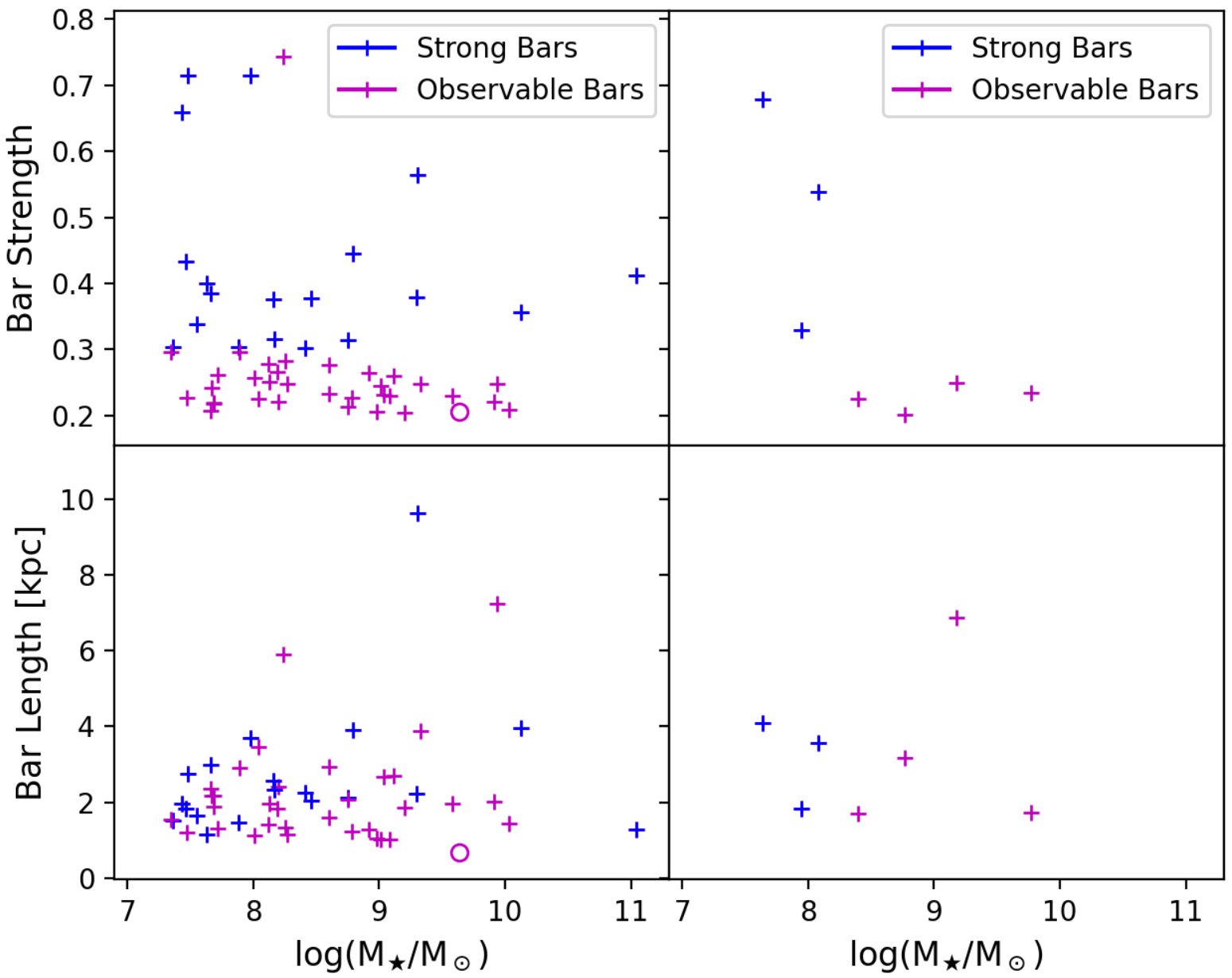}
    \caption{
    Figure showing bar strengths vs stellar mass (top panels) and bar lengths vs stellar mass (bottom panels) for $z=$ 1.3 (left panels) and $z=$ 0.25 (right panels). Bar strengths of both observable and strong bars are plotted for bars found in disc-dominated galaxies by the Velocity Harmonic Decomposition method. Bars of lengths greater than 1 kpc are marked with a cross and `short bars’ (those with lengths between 0.5 and 1.0 kpc) are marked as open circles. 
    One can see that the most strongly barred galaxies tend to be those of the lowest masses whereas the higher mass galaxies tend to host weaker bars - similar to the surface density method (Figure \protect\ref{fig:Bar_Strength_Vs_Mass}).    }
    \label{fig:Bar_Strength_Vs_Mass_Velocity}
\end{figure}

Figure \ref{fig:Bar_Strength_Vs_Mass_Velocity} presents plots of both bar strength and bar length vs stellar mass as calculated by the velocity method for comparison with the corresponding plot from the surface density method (Figure \ref{fig:Bar_Strength_Vs_Mass}). From this, we find the same trends between the two methods, that being that we find a slight anti-correlation between bar strengths and galaxy mass at \(z\) = 0.25 only and no correlation for bar lengths. We also find that there is almost no evolution in this relation between the bar strength and galaxy mass with redshift and that the higher mass galaxies tend to hold longer bars. By comparing between the two methods, we confirm a potentially barred galaxy as such. However, this does highlight biases in each method whereby these methods produce false positives and can incorrectly identify a galaxy as barred when used on their own. As discussed in the text, this could lead to current or future bar studies being biased to certain bar-like features without the incorporation of a secondary detection method.

\section{Linear stability of toy model with varying  budge \& halo mass}\label{sec:Toymodel}

In this appendix we briefly revisit the seminal  toy model of \cite{Aoki1979}  to compute quasi-analytically the fastest  growing modes of a  Toomre-Kuzmin disc, as a proxy for the onset of bar formation. For the purpose of this paper, we  extend it by accounting for the relative mass fraction in a  passive halo and bulge, while keeping the total disc-plus-bulge mass constant, allowing for mass transfer from the disc to the budge on the one hand, and from the cosmic web to the galactic disc on the other hand. This  provides us with a qualitative proxy for the trend found in the main text, i.e. that both reduced  disc mass and bulge growth quench bar formation. We rely on two significant (simplicity driven) caveats: i) the disc is assumed to be razor thin, gaseous, isolated; ii) we  limit our analysis to the computation of the linear growth rates of \textit{global} bi-symmetric perturbations \citep[i.e. beyond a local radial criterion such as Toomre's $Q$ number][]{Toomre_1963}. 

While the model for describing the response of the individual entities is crude, the joint analysis of  the  ensemble of timelines reflecting diverse environments should  provides us  with a more accurate global picture. The derived growth rate are an alternative  diagnostic for the dynamical state of  the  galaxies in the simulation. The assumption here is that their secular evolution with cosmic time can be approximated as a sequence of parameters in this model, so that, as the relative mass fractions and scale lengths change, the likelihood of a bi-symmetric instability developing into a strong bar varies.

\subsection{Toy model setup}

Consider a self-gravitating thin disc with a central spherical bulge and a spherical dark halo, 
with surface density $\Sigma$, and gravitational potential $\Phi$. For simplicity we will model this  disc
 as though it was made of gas.
Let $\boldsymbol{v}$ be its velocity field and $P$ its pressure. Then it obeys the hydro-dynamical and mean field equations
\begin{subequations}
\begin{align}
&\frac{\partial \Sigma}{\partial t} + \boldsymbol{\nabla} \cdot (\Sigma \boldsymbol{v}) = 0 \,, \label{eq:defsystem1} \\
&\frac{\partial \boldsymbol{v}}{\partial t} + (\boldsymbol{v} \cdot \boldsymbol{\nabla})\boldsymbol{v} = -\frac{1}{\Sigma} \boldsymbol{\nabla} P - \boldsymbol{\nabla} \Phi \,,\label{eq:defsystem2} \\
& \Delta \Phi_{\mathrm{disc}} = 4\pi G \Sigma \delta_\mathrm{D} (z), \label{eq:defsystem3}
\end{align}
\end{subequations}

where the total potential obeys $\Phi =\Phi_{\mathrm{bulge}}+  \Phi_{\mathrm{halo}}+  \Phi_{\mathrm{disk}} $. 
Let us assume the gas is a  polytrope with index $\Gamma \!=\! 4/3$, and fix the baryonic mass of the disc plus bulge to be $M=M_{\mathrm{disk}}+M_{\mathrm{bulge}}$. Then the system is described by the two parameters $p\!=\!M_{\mathrm{bulge}}/(M_{\mathrm{disk}}+M_{\mathrm{bulge}})$, the bulge fraction, and ${q\!=\!M_{\mathrm{disk}}/(M_{\mathrm{disk}}+M_{\mathrm{halo}})}$, the disc fraction.

The equilibrium state is modelled for simplicity by  Plummer spheres (for the bulge and halo) and a Kuzmin-Toomre  disc as
\begin{equation}
\Phi_{\mathrm{disc}} =-\frac{G M (1-p)}{\ad } \bigg(\frac{1-\xi}{2}\bigg)^{1/2} \,,
\end{equation}
and
\begin{equation}
\Phi_{\mathrm{bulge}}+\Phi_{\mathrm{halo}}=-\frac{G M }{\ad  } \bigg[\frac{\ad}{\ab}\frac{p}{\sqrt{1+(r/\ab)^2}}+\frac{\ad}{\ah}\bigg(\frac{1}{q}-1\bigg)\frac{1-p}{\sqrt{1+(r/\ah)^2}}\bigg]\,, \notag
\end{equation}
where $\ad$ is the  scale length of the disk, $\ah$ is the  scale length of the halo, $\ab$ that of the bulge, and where we introduced the reduced radius,   ${\xi = (r^2-\ad^2)/(r^2+\ad^2) }$. Then the angular velocity, $\Omega$, and the epicyclic frequency,  $\kappa$, of the whole system  read 
 \begin{align}
&\Omega(\xi) =\sqrt{\frac{GM(1-p)}{\ad^3}}  \bigg[\bigg( \frac{\ad}{\ab}\bigg)^3\frac{p}{1-p}  \bigg(\frac{1}{1+(r/\ab)^2}\bigg)^{3/2} \\
&\hskip -0.2cm+\bigg( \frac{\ad}{\ah}\bigg)^3\bigg(\frac{1}{q}-1\bigg)  \bigg(\frac{1}{1+(r/\ah)^2}\bigg)^{3/2} 
+   \bigg(\frac{1-\xi}{2}\bigg)^{3/2} \bigg(1-  \frac{\varepsilon_0}{ (1-p)^{2/3}} \bigg) \bigg]^{1/2} \notag,
\end{align}
\begin{equation}
    \frac{\kappa^2(\xi)}{2\Omega(\xi)}  = 2\Omega(\xi) \bigg[1+\frac{(1+\xi)(1-\xi)}{2\Omega}\frac{\mathrm{d} \Omega }{\mathrm{d} \xi} \bigg]\, ,
\end{equation}
where we introduced  $\varepsilon_0 = 0.1$ as the ratio of the internal and total energy 
which accounts for the strength of the  pressure forces in the disc \citep[given our choice of $\Gamma$ polytropic index, see][]{Aoki1979}. 

Let us  linearise the system~\eqref{eq:defsystem1}-\eqref{eq:defsystem2} in polar coordinates $(r,\theta)$, assuming an angular- and  time-dependant scaling in $\exp(\imath m \theta -\imath \omega t)$, and expand the first order perturbation
of the two components of Euler's equation and Poisson's equation over normalized Legendre polynomial in $\xi$ as, e.g. 
\begin{equation}
\Sigma^1(r,\theta, t) = \frac{M(1-p)}{2\pi \ad^2} \bigg(\frac{1-\xi}{2}\bigg)^{3/2}\!\!\! \sum_{n=|m|}^{\infty} \anm \Pnm(\xi) e^{\imath m \theta -\imath \omega t}\,, \notag
\end{equation}
and a similar expression for the radial and the azimuthal component of the perturbed velocity field involving 
$b_n^m$ and $c_n^m$ coefficients. By design, this expansion satisfies Poisson's equation.
Following closely \cite{Aoki1979}, upon injecting these expansions in the linearised system, multiplication by $ \Plm(\xi)$
and integration over $\xi$, using the orthogonality relation (for all $m$)
\begin{equation}
\int_{-1}^{1} \rd \xi \Pnm(\xi)\Plm(\xi) = \delta_{nl},
\end{equation}
 we obtain the following eigen-problem for the growth rate $\omega$ and the shape of the mode $\anm$:
\begin{subequations}
\begin{align}
&\sum_{n=|m|}^{\infty}  A_{ln} \anm  +\sum_{n=|m|}^{\infty}  B_{ln}\bnm+\sum_{n=|m|}^{\infty}  C_{ln}\cnm= \lambda \alm  ,  \label{eq:defmat1}\\
&\sum_{n=|m|}^{\infty}D_{ln}\anm  + \sum_{n=|m|}^{\infty}   A_{ln} \bnm+  \sum_{n=|m|}^{\infty} F_{ln} \cnm  =  \lambda    \blm ,
\\
&\sum_{n=|m|}^{\infty}G_{ln}\anm  + \sum_{n=|m|}^{\infty}   H_{ln} \bnm+  \sum_{n=|m|}^{\infty} A_{ln} \cnm  =  \lambda    \clm , \label{eq:defmat3}
 \end{align}
 \end{subequations}
where $\lambda= \mathrm{sign}(m)\,\omega /  \sqrt{GM/a^3}$ and with the matrix elements,
\begin{equation}
A_{ln} = |m| \sqrt{\frac{\ad^3}{GM}} \int_{-1}^{1} \rd \xi  \Plm(\xi)\Omega(\xi)\Pnm(\xi)\, , \notag
\end{equation}
\begin{equation}
B_{ln} = 4 \sqrt{1-p} \int_{-1}^{1} \rd \xi  \Plm(\xi)\bigg(\frac{1-\xi}{2}\bigg)^{1/2}\frac{\rd}{\rd \xi} \bigg[\bigg(\frac{1-\xi}{2}\bigg)^{5/4}\Pnm(\xi)\bigg]\, , \notag
\end{equation}
\begin{equation}
C_{ln} = |m| \sqrt{1-p} \int_{-1}^{1} \rd \xi  \Plm(\xi)\bigg(\frac{1-\xi}{2}\bigg)^{3/4}\bigg(\frac{1+\xi}{2}\bigg)^{-1}\Pnm(\xi)\, , \notag
\end{equation}
\begin{align}
D_{ln} &= 4 \sqrt{1-p} \bigg(\frac{1}{2n+1}-\frac{\varepsilon_0}{3(1-x)^{2/3}}\bigg)\notag\\
&\times \int_{-1}^{1} \rd \xi  \Plm(\xi)\bigg(\frac{1-\xi}{2}\bigg)^{5/4}\bigg(\frac{1+\xi}{2}\bigg)\frac{\rd}{\rd \xi} \bigg[\bigg(\frac{1-\xi}{2}\bigg)^{1/2}\Pnm(\xi)\bigg]\, \notag, 
\end{align}
\begin{equation}
G_{ln} \!=\! -|m| \sqrt{1-p} \bigg(\frac{1}{2n\!+\!1}-\frac{\varepsilon_0}{3(1\!-\!x)^{2/3}}\bigg) 
\!\! \int_{-1}^{1}\!\!\! \rd \xi  \Plm(\xi)\bigg(\frac{1-\xi}{2}\bigg)^{3/4}\!\!\!\! \Pnm(\xi)\, \notag, 
\end{equation}
\begin{equation}
H_{ln} = \sqrt{\frac{\ad^3}{GM}} \int_{-1}^{1} \rd \xi  \Plm(\xi)\frac{\kappa^2(\xi)}{2\Omega(\xi)}\Pnm(\xi)\,,\notag
\end{equation}
and $F_{ln}=2/|m| A_{ln}$.
The integrals involving $\Omega$ and $\kappa$ are computed using a mid-point sampling rule, while the others are computed analytically by recursion.

Finding the growth rate of the mode, $\omega$, involves the computation of the eigenvalues of the infinite response matrix defined by Equations~\eqref{eq:defmat1}-\eqref{eq:defmat3}.
 In practice, we only have access to  sequence of  truncated matrices of varying size set by  $n_\mathrm{max}$, from which we obtain lists of eigenvalues. Among those, we  determine which  are physically relevant, and  which result from the truncation process. The latter either diverge to infinity or oscillate, which allows us to disregard them in favour of the subset which converges with $n_\mathrm{max}$ (here typically $n_\mathrm{max}\leq 170$).
 For each $p,q$ we then select that with highest imaginary part, $\omega_{\mathrm{I}} = \mathrm{Im}(\omega)$, and build the corresponding map, within which we can assign some threshold below which we consider the $m-$fold symmetric mode grows too slowly for bar formation. 
In practice, the fastest growing mode shape and pattern speed displays sets of discontinuities as one increases $p$ (or decreases $q$), as eigenvalues corresponding to distinct physical branches become dominant. This can be seen as wiggles in the bottom left part of the light contours on  figure~\ref{fig:disc-stab}.

Eventually, it could be of interest and more realistic
to implement the fitting strategy presented in \cite{Ueda1985} to our sets of galaxies, 
while accounting for the detailed shape of the rotation curves (bulge included) and surface densities, in order to quantify statistically the 
dynamical stability of \newh discs.
Another possible improvement would involve implementing a proper stellar stability analysis, following e.g. \cite{Pichon_1997},
which provides the flexibility in matching the distribution function and potential to that of the simulated galaxies. One could eventually account for the disc's thickness, the live halo, or use the shape of the eigenvectors to match the pitch angle of the spiral  response.

\subsection{Mapping \newh galaxies onto the \((p,q)\) plane} \label{sec:mapping}
To interpret the \newh galaxies in the context of Figure~\ref{fig:disc-stab}, we require a mapping of the galaxies onto \((p,q)\) space. In the following mapping, we consider galaxies in the \(z=1.3\) snapshot, but the mappings are similar for other snapshots. We measure the best mapping as follows. First, we fit the rotation curve of the dark matter halo, \(v_{c,{\rm halo}}\) using the rotation curve of a Plummer sphere. We define the log likelihood function for the rotation curve evaluated at \(i\) radial points \(r_i\) in the galaxy midplane, with associated Poisson-like uncertainty \(\sigma_i\), given the free parameters for the Plummer sphere representing the halo (\(a_{\rm h}\), the scale length of the halo, and \(M_{\rm halo}\), the mass of halo) as 
\begin{equation}
    \ln \mathcal{L}\left(v_c|a_{\rm h},M_{\rm halo}\right) = -\frac{1}{2}\sum_i \left[\frac{\left(v_{c,i} - v_{c,m}(r_i)\right)^2}{\sigma_i^2} \right]\,, \label{eq:rotcurve_likelihood}
\end{equation}
where the model rotation curve at each radial point \(r_i\) is given by
\begin{equation}
    v_{c,m}(r_i) = \sqrt{2{GM}_{\rm halo}r_i^2(a_{\rm h}^2 + r_i^2)^{-(3/2)}}. \label{eq:rotcurve_model}
\end{equation}

The uncertainty at each point on the rotation curve (\(\sigma_i\)) is attributable to several factors, such as enclosed particle number, centre choice, and angular momentum rectification. Of these, the enclosed particle number contributes the most to the uncertainty budget. We therefore estimate the uncertainty on each point in the rotation curve using the reciprocal of the enclosed particle number at each radius, defining the minimum uncertainty to be 3\%. In practice, the fits exhibit little dependence on the choice of uncertainty, such that a fixed uncertainty returns nearly the same parameter estimates. We then use {\sc emcee} \citep{ForemanMackey_2013} to perform a Markov Chain Monte Carlo estimation of \(a_{\rm h}\) and \(M_{\rm halo}\). Comparing the \(M_{\rm halo}\) estimates to the true measured halo masses, we find a consistent bias to lower halo masses across the galaxy mass range related to the mismatch of the rotation curve profiles. The bias is constant with halo mass, and fixed at \(\log(M_{\rm halo,fit})-\log(M_{\rm halo,true})=-0.3\). We correct for the bias below when estimating \(q\). In addition to the bias, we also find modest variance (\(\pm0.1\) dex) in the fit values, which contributes to uncertainty in the calculation of \(q\).

Second, we fit a Toomre-Kuzmin rotation curve profile to the stellar disc component of the galaxies. Conveniently, the Plummer sphere and Toomre-Kuzmin disc have the same rotation curve profile in the \(z=0\) plane, where we are performing the fits. Therefore, in equations \ref{eq:rotcurve_likelihood} and \ref{eq:rotcurve_model}, we need only change the subscripts to reflect that we are now estimating the disc parameters \(a_{\rm d}\) and \(M_{\rm disc}\). We again use the same uncertainty estimation scheme, and estimate the parameters using {\sc emcee}. We then compute the residuals for each disc galaxy. We define a galaxy as well-described by a single Toomre-Kuzmin disc component if the maximum relative error \((v_{c,i} - v_{c,m}(r_i))/v_{c,m}(r_i)<0.1\). We find that \(>90\%\) of galaxies (291 of 299 galaxies at \(z=1.3\)) meet this criteria. 

We again compare the fit masses with the measured masses, finding similar bias and variance as in the halo (i.e. with no dependence on galaxy mass). We correct for the bias below when computing \(q\). We may also compare the fit \(a_{\rm d}\) to the measured \(r_{v_{\star,{\rm max}}}\) in Section~\ref{subsection:rotationcurves}, finding that \(\langle a_{\rm d}/r_{v_{\star,{\rm max}}}\rangle=0.6\). For the galaxies that are well-fit by a Toomre-Kuzmin disc and Plummer halo, we define \(\log q=\log(M_{\rm disc}/(M_{\rm halo}+M_{\rm disc}))\) and \(p=0\). We find, after correcting for the bias in halo mass, that the galaxies are well-fit by the relation \(\log q_{z=1.3}=0.6\log(M_{\rm halo,true})-7.5\), resulting in \(\langle\log q \rangle=-1.7^{+0.3}_{-0.4}\) for a `typical' galaxy in the sample at \(\log(M_{\rm halo})=9.7\). We therefore define \((p,\log q)=(0.0,-1.7)\) as the location in \((p,\log q)\) space for a typical bulgeless galaxy. From the fits to the galaxies, we find that \(\langle a_{\rm h}/a_{\rm d}\rangle = 2.8\pm0.3\), which we use to inform the toy model construction. This ratio is valid for all galaxies, whether or not they are bulge-hosting.

Through visual inspection, we define galaxies which have maximum relative error \(>0.1\) as `bulge-hosting' galaxies, which require a second component. The galaxies that require a second component are those at the high mass end, as expected from the rotation curve decomposition in Section~\ref{subsection:rotationcurves} and Figure~\ref{fig:summarystatistics}. These galaxies also have larger \(\log q\) values on average, following the trend line listed above, such that if \(p=0\), they may fall in the instability regime. We estimate the bulge component for these galaxies using a two-component model for the rotation curve, 
\begin{equation}
 v_{c,m}(r_i) = \bigg[2Gr_i^2\bigg(\frac{M_{\rm disc}}{(a_{\rm d}^2 + r_i^2)^{3/2}}
    +\frac{M_{\rm bulge}}{(a_{\rm b}^2 + r_i^2)^{3/2}}\bigg)\bigg]^{1/2}\,,
\end{equation}
where we require that \(a_{\rm b}<a_{\rm d}\) and \(M_{\rm bulge}<M_{\rm disc}\). These criteria enforce the disc as the dominant mass component, and require that the bulge be compact relative to the disc, consistent with what is observed in the galaxies by visual inspection. The rest of the likelihood function is unchanged apart from the addition of the dependence of the data on two extra parameters. From the fits to these galaxies, we find that \(\langle a_{\rm d}/a_{\rm b}\rangle = 20\pm4\), which we use to inform the toy model construction. In practice, we then end up with `maximal' disc components (i.e. contributions from the Toomre-Kuzmin disc that match the rotation curve when \(r\) is large). Also from the fits, we compute the \(p\) values to be \(p\approx 0.2\). We therefore define \((p,\log q)=(0.2,-1.0)\) as the location in \((p,\log q)\) space for a typical bulge-hosting galaxy.

We also perform a cursory inspection of galaxies at \(z=0.25\), the lowest redshift analysed in our sample. The parameters of the toy model remain largely unchanged, with a modest increase in \(a_{\rm d}/a_{\rm b}\) (i.e. bulges have become more concentrated), which has the effect of moving the stability line towards increasing \(x\). We find that on average, \(\log q_{z=0.25}=\log q_{z=1.3}+0.2\). A significant fraction of the disc galaxies still do not appear to host a bulge, while the typical bulge-hosting galaxy has increased the location in \(p\)-space to 0.25. Taken together, the disc-dominated galaxies appear to be growing self-similarly while the bulges grow at a modestly faster rate and with decreasing scale length, which serves to keep the discs stable against bar formation.

\section{High Mass Bar Fractions}

In this section, we present the results of our analysis in terms of bar fractions when placing a stellar mass cut of $\textup{M}_{\star} \geq 10^{10} \textup{M}_{\odot}$ so as to provide a more versatile comparison to observations in the literature.

\begin{table*}
\caption{The results of our analysis in terms of spiral fractions and bar fractions for 'strong bars' ($S \geq 0.3$) and 'observable bars' ($S \geq 0.2$) at redshifts \(z = 1.3, 1.0, 0.7, 0.5\) and \(0.25\). The bar fractions as obtained by both bar detection methods described in Section \ref{bar_analysis} are shown. The confidence intervals presented have been estimated using the (Bayesian) beta distribution quantile technique \citep{Cameron_2011}. We also present numerical values in the parenthesis.
We have applied a mass cut $\textup{M}_{\star} \geq 10^{10} \textup{M}_{\odot}$ so as to compare more directly to previous observational studies. See Figure \ref{tab:Bar_Fractions} for our full results.}
\centering
\begin{tabular}{ccccc}
\cline{4-5}
                      & \multicolumn{1}{l}{} &                 & \multicolumn{2}{c}{\begin{tabular}[c]{@{}c@{}}Bar Fractions\\ (Number of Bars)\end{tabular}}                                                                       \\ \hline
Redshift              & \begin{tabular}[c]{@{}c@{}}Spiral Fraction\\ (Number of Spirals)\end{tabular} & Method          & Strong Bars                                          & Observable Bars                                             \\ \hline
\multirow{3}{*}{1.3}  & \multirow{3}{*}{\begin{tabular}[c]{@{}c@{}}$0.923_{-0.140}^{+0.026}$\\ (12)\end{tabular}} & Surface Density & \begin{tabular}[c]{@{}c@{}}$0.167_{-0.059}^{+0.155}$\\ (2)\end{tabular}  & \begin{tabular}[c]{@{}c@{}}$0.250_{-0.083}^{+0.156}$\\ (3)\end{tabular}  \\  
                      &                       & Velocity        & \begin{tabular}[c]{@{}c@{}}$0.167_{-0.059}^{+0.155}$\\ (2)\end{tabular} & \begin{tabular}[c]{@{}c@{}}$0.250_{-0.083}^{+0.156}$\\ (3)\end{tabular} \\ \hline
\multirow{3}{*}{1.0}  & \multirow{3}{*}{\begin{tabular}[c]{@{}c@{}}$1.000_{-0.103}^{+0.010}$\\ (16)\end{tabular}} & Surface Density & \begin{tabular}[c]{@{}c@{}}$0.063_{-0.020}^{+0.119}$\\ (1)\end{tabular}  & \begin{tabular}[c]{@{}c@{}}$0.063_{-0.020}^{+0.119}$\\ (1)\end{tabular}  \\ 
                      &                       & Velocity        & \begin{tabular}[c]{@{}c@{}}$0.063_{-0.020}^{+0.119}$\\ (1)\end{tabular}  & \begin{tabular}[c]{@{}c@{}}$0.063_{-0.020}^{+0.119}$\\ (1)\end{tabular}  \\ \hline
\multirow{3}{*}{0.7}   & \multirow{3}{*}{\begin{tabular}[c]{@{}c@{}}$0.857_{-0.140}^{+0.050}$\\ (12)\end{tabular}} & Surface Density & \begin{tabular}[c]{@{}c@{}}$0.0$\\ (0)\end{tabular}  & \begin{tabular}[c]{@{}c@{}}$0.083_{-0.028}^{+0.149}$\\ (1)\end{tabular}   \\ 
                      &                      & Velocity        & \begin{tabular}[c]{@{}c@{}}$0.0$\\ (0)\end{tabular}  & \begin{tabular}[c]{@{}c@{}}$0.083_{-0.028}^{+0.149}$\\ (1)\end{tabular}   \\ \hline
\multirow{3}{*}{0.5}  & \multirow{3}{*}{\begin{tabular}[c]{@{}c@{}}$0.765_{-0.128}^{+0.072}$\\ (13)\end{tabular}} & Surface Density & \begin{tabular}[c]{@{}c@{}}$0.077_{-0.026}^{+0.140}$\\ (1)\end{tabular}  & \begin{tabular}[c]{@{}c@{}}$0.077_{-0.026}^{+0.140}$\\ (1)\end{tabular}   \\ 
                      &                      & Velocity        & \begin{tabular}[c]{@{}c@{}}$0.077_{-0.026}^{+0.140}$\\ (1)\end{tabular}  & \begin{tabular}[c]{@{}c@{}}$0.154_{-0.054}^{+0.147}$\\ (2)\end{tabular}   \\ \hline
\multirow{3}{*}{0.25} & \multirow{3}{*}{\begin{tabular}[c]{@{}c@{}}$0.714_{-0.114}^{+0.077}$\\ (15)\end{tabular}} & Surface Density & \begin{tabular}[c]{@{}c@{}}$0.0$\\ (0)\end{tabular}  & \begin{tabular}[c]{@{}c@{}}$0.0$\\ (0)\end{tabular}   \\ 
                      &                      & Velocity        & \begin{tabular}[c]{@{}c@{}}$0.0$\\ (0)\end{tabular}  & \begin{tabular}[c]{@{}c@{}}$0.0$\\ (0)\end{tabular}  \\ \hline
\end{tabular}
\label{tab:Bar_Fractions_Mass_Cut}
\end{table*}

\begin{table}
\caption{The final results of our bar search analysis in terms of bar fractions calibrated between the two methods whereby a barred galaxy is confirmed as such if said galaxy found to be barred in by both methods. The bar fractions (and number of barred galaxies) are shown for redshifts \(z = 1.3, 1.0, 0.7, 0.5\) and \(0.25\). The confidence intervals presented have been estimated using the (Bayesian) beta distribution quantile technique \citep{Cameron_2011}.
We have applied a mass cut $\textup{M}_{\star} \geq 10^{10} \textup{M}_{\odot}$ so as to compare more directly to previous observational studies. See Figure \ref{tab:Final_Bar_Fractions} for our full results.}
\centering
\begin{tabular}{ccc}
\cline{2-3}
         & \multicolumn{2}{c}{\begin{tabular}[c]{@{}c@{}}Final Bar Fractions\\ (Number of Bars)\end{tabular}}                                                                \\ \hline
Redshift & Strong Bars                                         & Observable Bars                                                                      \\ \hline
1.3      & \begin{tabular}[c]{@{}c@{}}0.167$_{-0.059}^{+0.155}$\\ (2)\end{tabular} & \begin{tabular}[c]{@{}c@{}}0.167$_{-0.059}^{+0.155}$\\ (2)\end{tabular}  \\
1.0      & \begin{tabular}[c]{@{}c@{}}0.0\\ (0)\end{tabular} & \begin{tabular}[c]{@{}c@{}}0.063$_{-0.020}^{+0.119}$\\ (1)\end{tabular}  \\
0.7      & \begin{tabular}[c]{@{}c@{}}0.0\\ (0)\end{tabular} & \begin{tabular}[c]{@{}c@{}}0.083$_{-0.028}^{+0.149}$\\ (1)\end{tabular}  \\
0.5      & \begin{tabular}[c]{@{}c@{}}0.077$_{-0.026}^{+0.140}$\\ (1)\end{tabular} & \begin{tabular}[c]{@{}c@{}}0.077$_{-0.026}^{+0.140}$\\ (1)\end{tabular}  \\
0.25     & \begin{tabular}[c]{@{}c@{}}0.0\\ (0)\end{tabular} & \begin{tabular}[c]{@{}c@{}}0.0\\ (0)\end{tabular}  
\end{tabular}
\label{tab:Final_Bar_Fractions_Mass_Cut}
\end{table}

\label{lastpage}
\end{document}